\documentclass[ reprint, superscriptaddress, amsmath,amssymb, aps]{revtex4-1}
\usepackage{graphicx}
\usepackage{dcolumn}
\usepackage{amsthm}
\usepackage{bm}
\newcommand{\ket}[1]{|#1\rangle}

\pdfoutput=1
\usepackage{graphicx,graphics,epsfig,subfigure,times,bm,bbm,amssymb,amsmath,amsthm,mathrsfs,MnSymbol}
\usepackage{gensymb}
\usepackage{amsfonts}
\usepackage[matrix,frame,arrow]{xypic}
\usepackage[pdfstartview=FitH]{hyperref}
\hypersetup{
	colorlinks=true,       
	linkcolor=red,          
	citecolor=magenta,        
	filecolor=magenta,      
	urlcolor=cyan,           
	runcolor=cyan}
\usepackage{epstopdf}
\usepackage[pdftex]{color}
\usepackage{braket}
\usepackage{enumerate}
\usepackage[normalem]{ulem}
\usepackage[usenames,dvipsnames]{xcolor}
\usepackage{multirow}
\usepackage{mathtools}

\definecolor{orange}{rgb}{1,0.5,0}

\newcommand{\ignore}[1]{}
\usepackage{geometry}

\begin{document}

\title{Observation of thermalization and information scrambling in a superconducting quantum processor}

\author{Qingling Zhu}
\thanks{Those authors contributed equally to this work.}
\affiliation{Hefei National Laboratory for Physical Sciences at Microscale and Department of Modern Physics, University of Science and Technology of China, Hefei, Anhui 230026, China}
\affiliation{Shanghai Branch, CAS Center for Excellence and Synergetic Innovation Center in Quantum Information and Quantum Physics, University of Science and Technology of China, Shanghai 201315, China}
\affiliation{Shanghai Research Center for Quantum Sciences, Shanghai 201315, China}

\author{Zheng-Hang Sun}
\thanks{Those authors contributed equally to this work.}
\affiliation{Institute of Physics, Chinese Academy of Sciences, Beijing 100190, China}
\affiliation{School of Physical Sciences, University of Chinese Academy of Sciences, Beijing 100190, China}

\author{Ming Gong}
\thanks{Those authors contributed equally to this work.}
\affiliation{Hefei National Laboratory for Physical Sciences at Microscale and Department of Modern Physics, University of Science and Technology of China, Hefei, Anhui 230026, China}
\affiliation{Shanghai Branch, CAS Center for Excellence and Synergetic Innovation Center in Quantum Information and Quantum Physics, University of Science and Technology of China, Shanghai 201315, China}
\affiliation{Shanghai Research Center for Quantum Sciences, Shanghai 201315, China}

\author{Fusheng Chen}
\affiliation{Hefei National Laboratory for Physical Sciences at Microscale and Department of Modern Physics, University of Science and Technology of China, Hefei, Anhui 230026, China}
\affiliation{Shanghai Branch, CAS Center for Excellence and Synergetic Innovation Center in Quantum Information and Quantum Physics, University of Science and Technology of China, Shanghai 201315, China}
\affiliation{Shanghai Research Center for Quantum Sciences, Shanghai 201315, China}

\author{Yu-Ran Zhang}
\affiliation{Theoretical Quantum Physics Laboratory, RIKEN Cluster for Pioneering Research, Wako-shi, Saitama 351-0198, Japan}

\author{Yulin Wu}
\affiliation{Hefei National Laboratory for Physical Sciences at Microscale and Department of Modern Physics, University of Science and Technology of China, Hefei, Anhui 230026, China}
\affiliation{Shanghai Branch, CAS Center for Excellence and Synergetic Innovation Center in Quantum Information and Quantum Physics, University of Science and Technology of China, Shanghai 201315, China}
\affiliation{Shanghai Research Center for Quantum Sciences, Shanghai 201315, China}

\author{Yangsen Ye}
\affiliation{Hefei National Laboratory for Physical Sciences at Microscale and Department of Modern Physics, University of Science and Technology of China, Hefei, Anhui 230026, China}
\affiliation{Shanghai Branch, CAS Center for Excellence and Synergetic Innovation Center in Quantum Information and Quantum Physics, University of Science and Technology of China, Shanghai 201315, China}
\affiliation{Shanghai Research Center for Quantum Sciences, Shanghai 201315, China}

\author{Chen Zha}
\affiliation{Hefei National Laboratory for Physical Sciences at Microscale and Department of Modern Physics, University of Science and Technology of China, Hefei, Anhui 230026, China}
\affiliation{Shanghai Branch, CAS Center for Excellence and Synergetic Innovation Center in Quantum Information and Quantum Physics, University of Science and Technology of China, Shanghai 201315, China}
\affiliation{Shanghai Research Center for Quantum Sciences, Shanghai 201315, China}

\author{Shaowei Li}
\affiliation{Hefei National Laboratory for Physical Sciences at Microscale and Department of Modern Physics, University of Science and Technology of China, Hefei, Anhui 230026, China}
\affiliation{Shanghai Branch, CAS Center for Excellence and Synergetic Innovation Center in Quantum Information and Quantum Physics, University of Science and Technology of China, Shanghai 201315, China}
\affiliation{Shanghai Research Center for Quantum Sciences, Shanghai 201315, China}

\author{Shaojun Guo}
\affiliation{Hefei National Laboratory for Physical Sciences at Microscale and Department of Modern Physics, University of Science and Technology of China, Hefei, Anhui 230026, China}
\affiliation{Shanghai Branch, CAS Center for Excellence and Synergetic Innovation Center in Quantum Information and Quantum Physics, University of Science and Technology of China, Shanghai 201315, China}
\affiliation{Shanghai Research Center for Quantum Sciences, Shanghai 201315, China}

\author{Haoran Qian}
\affiliation{Hefei National Laboratory for Physical Sciences at Microscale and Department of Modern Physics, University of Science and Technology of China, Hefei, Anhui 230026, China}
\affiliation{Shanghai Branch, CAS Center for Excellence and Synergetic Innovation Center in Quantum Information and Quantum Physics, University of Science and Technology of China, Shanghai 201315, China}
\affiliation{Shanghai Research Center for Quantum Sciences, Shanghai 201315, China}

\author{He-Liang Huang}
\affiliation{Hefei National Laboratory for Physical Sciences at Microscale and Department of Modern Physics, University of Science and Technology of China, Hefei, Anhui 230026, China}
\affiliation{Shanghai Branch, CAS Center for Excellence and Synergetic Innovation Center in Quantum Information and Quantum Physics, University of Science and Technology of China, Shanghai 201315, China}
\affiliation{Shanghai Research Center for Quantum Sciences, Shanghai 201315, China}

\author{Jiale Yu}
\affiliation{Hefei National Laboratory for Physical Sciences at Microscale and Department of Modern Physics, University of Science and Technology of China, Hefei, Anhui 230026, China}
\affiliation{Shanghai Branch, CAS Center for Excellence and Synergetic Innovation Center in Quantum Information and Quantum Physics, University of Science and Technology of China, Shanghai 201315, China}
\affiliation{Shanghai Research Center for Quantum Sciences, Shanghai 201315, China}

\author{Hui Deng}
\affiliation{Hefei National Laboratory for Physical Sciences at Microscale and Department of Modern Physics, University of Science and Technology of China, Hefei, Anhui 230026, China}
\affiliation{Shanghai Branch, CAS Center for Excellence and Synergetic Innovation Center in Quantum Information and Quantum Physics, University of Science and Technology of China, Shanghai 201315, China}
\affiliation{Shanghai Research Center for Quantum Sciences, Shanghai 201315, China}

\author{Hao Rong}
\affiliation{Hefei National Laboratory for Physical Sciences at Microscale and Department of Modern Physics, University of Science and Technology of China, Hefei, Anhui 230026, China}
\affiliation{Shanghai Branch, CAS Center for Excellence and Synergetic Innovation Center in Quantum Information and Quantum Physics, University of Science and Technology of China, Shanghai 201315, China}
\affiliation{Shanghai Research Center for Quantum Sciences, Shanghai 201315, China}

\author{Jin Lin}
\affiliation{Hefei National Laboratory for Physical Sciences at Microscale and Department of Modern Physics, University of Science and Technology of China, Hefei, Anhui 230026, China}
\affiliation{Shanghai Branch, CAS Center for Excellence and Synergetic Innovation Center in Quantum Information and Quantum Physics, University of Science and Technology of China, Shanghai 201315, China}
\affiliation{Shanghai Research Center for Quantum Sciences, Shanghai 201315, China}

\author{Yu Xu}
\affiliation{Hefei National Laboratory for Physical Sciences at Microscale and Department of Modern Physics, University of Science and Technology of China, Hefei, Anhui 230026, China}
\affiliation{Shanghai Branch, CAS Center for Excellence and Synergetic Innovation Center in Quantum Information and Quantum Physics, University of Science and Technology of China, Shanghai 201315, China}
\affiliation{Shanghai Research Center for Quantum Sciences, Shanghai 201315, China}

\author{Lihua Sun}
\affiliation{Hefei National Laboratory for Physical Sciences at Microscale and Department of Modern Physics, University of Science and Technology of China, Hefei, Anhui 230026, China}
\affiliation{Shanghai Branch, CAS Center for Excellence and Synergetic Innovation Center in Quantum Information and Quantum Physics, University of Science and Technology of China, Shanghai 201315, China}
\affiliation{Shanghai Research Center for Quantum Sciences, Shanghai 201315, China}

\author{Cheng Guo}
\affiliation{Hefei National Laboratory for Physical Sciences at Microscale and Department of Modern Physics, University of Science and Technology of China, Hefei, Anhui 230026, China}
\affiliation{Shanghai Branch, CAS Center for Excellence and Synergetic Innovation Center in Quantum Information and Quantum Physics, University of Science and Technology of China, Shanghai 201315, China}
\affiliation{Shanghai Research Center for Quantum Sciences, Shanghai 201315, China}

\author{Na Li}
\affiliation{Hefei National Laboratory for Physical Sciences at Microscale and Department of Modern Physics, University of Science and Technology of China, Hefei, Anhui 230026, China}
\affiliation{Shanghai Branch, CAS Center for Excellence and Synergetic Innovation Center in Quantum Information and Quantum Physics, University of Science and Technology of China, Shanghai 201315, China}
\affiliation{Shanghai Research Center for Quantum Sciences, Shanghai 201315, China}

\author{Futian Liang}
\affiliation{Hefei National Laboratory for Physical Sciences at Microscale and Department of Modern Physics, University of Science and Technology of China, Hefei, Anhui 230026, China}
\affiliation{Shanghai Branch, CAS Center for Excellence and Synergetic Innovation Center in Quantum Information and Quantum Physics, University of Science and Technology of China, Shanghai 201315, China}
\affiliation{Shanghai Research Center for Quantum Sciences, Shanghai 201315, China}

\author{Cheng-Zhi Peng}
\affiliation{Hefei National Laboratory for Physical Sciences at Microscale and Department of Modern Physics, University of Science and Technology of China, Hefei, Anhui 230026, China}
\affiliation{Shanghai Branch, CAS Center for Excellence and Synergetic Innovation Center in Quantum Information and Quantum Physics, University of Science and Technology of China, Shanghai 201315, China}
\affiliation{Shanghai Research Center for Quantum Sciences, Shanghai 201315, China}

\author{Heng Fan}
\affiliation{Institute of Physics, Chinese Academy of Sciences, Beijing 100190, China}
\affiliation{School of Physical Sciences, University of Chinese Academy of Sciences, Beijing 100190, China}
\affiliation{Songshan Lake  Materials Laboratory, Dongguan 523808, Guangdong, China}
\affiliation{CAS Center for Excellent in Topological Quantum Computation, University of Chinese Academy of Sciences, Beijing 100190, China}

\author{Xiaobo Zhu}
\affiliation{Hefei National Laboratory for Physical Sciences at Microscale and Department of Modern Physics, University of Science and Technology of China, Hefei, Anhui 230026, China}
\affiliation{Shanghai Branch, CAS Center for Excellence and Synergetic Innovation Center in Quantum Information and Quantum Physics, University of Science and Technology of China, Shanghai 201315, China}
\affiliation{Shanghai Research Center for Quantum Sciences, Shanghai 201315, China}

\author{Jian-Wei Pan}
\affiliation{Hefei National Laboratory for Physical Sciences at Microscale and Department of Modern Physics, University of Science and Technology of China, Hefei, Anhui 230026, China}
\affiliation{Shanghai Branch, CAS Center for Excellence and Synergetic Innovation Center in Quantum Information and Quantum Physics, University of Science and Technology of China, Shanghai 201315, China}
\affiliation{Shanghai Research Center for Quantum Sciences, Shanghai 201315, China}

\begin{abstract}
\noindent \textbf{Understanding various phenomena in non-equilibrium dynamics of closed quantum many-body systems, such as  quantum thermalization, information scrambling, and nonergodic dynamics, is a crucial for modern physics. Using a ladder-type superconducting quantum processor, we perform analog quantum simulations of both the $XX$ ladder and one-dimensional (1D) $XX$ model. By measuring the dynamics of local observables, entanglement entropy and tripartite mutual information, we signal quantum thermalization and information scrambling in the $XX$ ladder. In contrast, we show that the $XX$ chain, as free fermions on a 1D lattice, fails to thermalize, and local information does not scramble in the integrable channel. Our experiments reveal ergodicity and scrambling in the controllable qubit ladder, and opens the door to further investigations on the thermodynamics and chaos in quantum many-body systems.}
\end{abstract}
\pacs{Valid PACS appear here}
\maketitle

Whether the out-of-equilibrium dynamics of a quantum many-body system can present thermalization~\cite{thermalization_rigol,integrable1} and information scrambling~\cite{scrambling_xiaoliang} is a fundamental issue in statistical mechanics. The occurrence or absence of ergodicity and information scrambling depends on whether integrability is broken or not. A nonintegrable system thermalizes when it evolves, where the quenched state can be described by the Gibbs distribution~\cite{thermalization_mbl}. However, thermalization is absent in integrable systems due to infinitely many conserved quantities~\cite{integrable1,integrable2}. Similarly, information scrambling cannot occur in 1D free fermions as an integrable system, while a generic non-integrable system scrambles information~\cite{scrambling_xiaoliang,scrambling_prb}. Experiments on quantum thermalization have been demonstrated in cold atoms~\cite{cold_atom} and trapped ions~\cite{trapped_ion} with time-independent Hamiltonians, as well as periodic Floquet systems~\cite{Floquet1,Floquet2}. In addition, information scrambling can be identified by out-of-time-order correlators (OTOCs)~\cite{scrambling_xiaoliang,tele_2}, which have been directly measured using time-reversal operations~\cite{exp_OTOC1,exp_OTOC2}. Nevertheless, the experimental implementation of both integrable and non-integrable systems on the same quantum processor, where distinguishable characteristics of ergodicity and information scrambling can be observed, remains limited.

\begin{figure}[]
	\centering
	\includegraphics[width=1\linewidth]{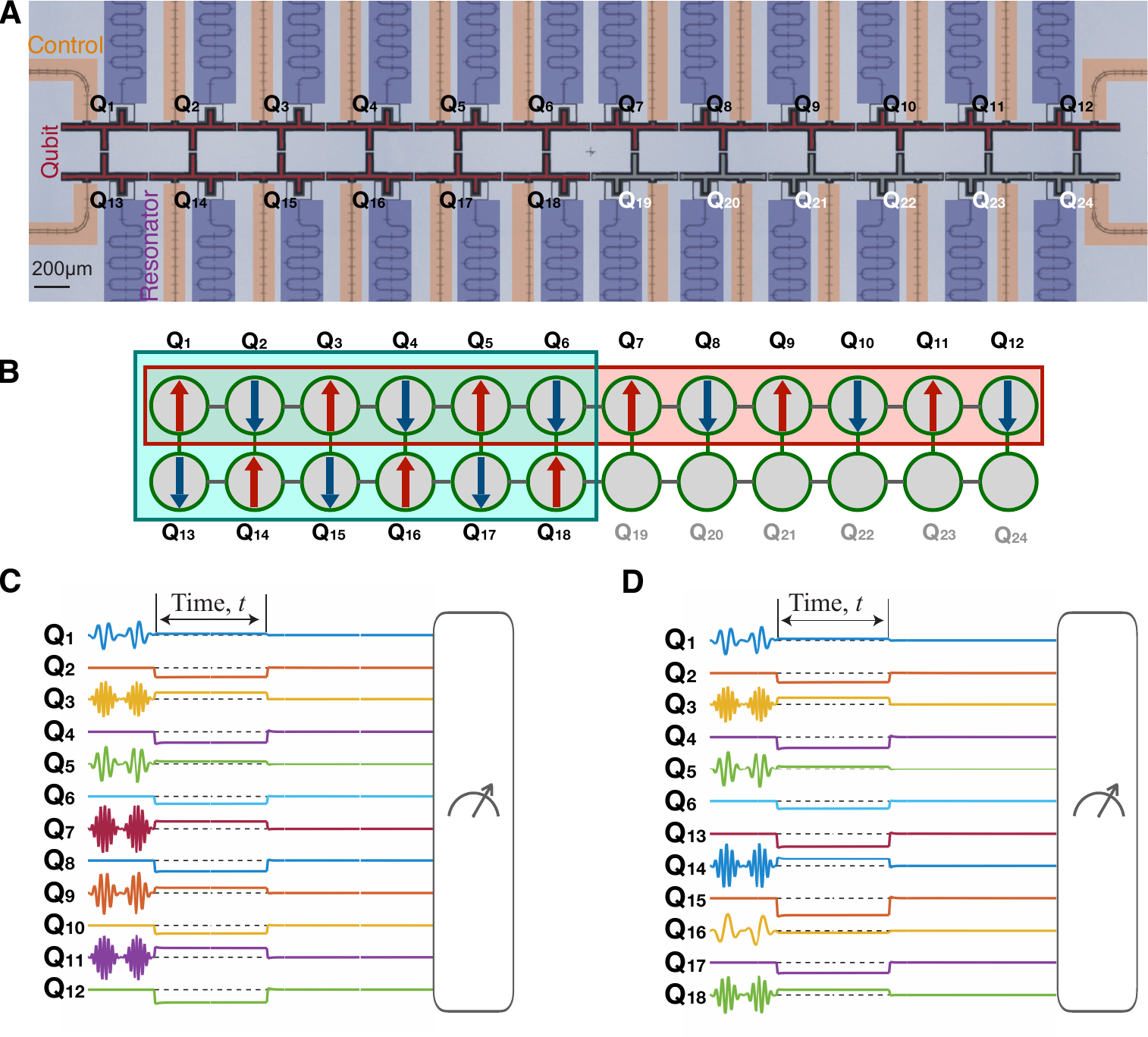}\\
	\caption{\textbf{Superconducting quantum circuit and experimental pulse sequences.} (\textbf{A}) False-color optical micrograph of the superconducting circuit. Each qubit has an independent control line for the XY and Z control (the yellow region), coupled to a separate readout resonator (the purple region). (\textbf{B}) A schematic graph of the superconducting quantum circuit. The up and down arrows indicate that the initial state of the qubit is $|1\rangle$ and $|0\rangle$, respectively. The qubits $Q_{1}$--$Q_{12}$ are employed to the quantum simulation of $XX$ chain. The qubits $Q_{1}$--$Q_{6}$ and $Q_{13}$--$Q_{18}$ are employed to the quantum simulation of $XX$ ladder. (\textbf{C}) and (\textbf{D}) are the experimental pulse sequence of the quantum simulation of $XX$ chain and ladder, respectively. The pulse sequences consist of initialization, evolution and readout. In the initialization, all qubits are at $|0\rangle$, and the X gates are applied on the qubits whose chosen initial state is $|1\rangle$. Next, the qubits are tuned to the working point via Z pulses, and the time evolution is realized. Finally, the measurements are performed after tuning the qubits back to their idle points.    }\label{exp_setup}
\end{figure}

Recent numerical works have shown that ergodicity and scrambling can occur in the $XX$ ladder~\cite{XX_ladder1,XX_ladder2}, but the 1D $XX$ model is a typical integrable system~\cite{XY_chain} that exhibits the characteristics of free fermions. Here, we realize
the $XX$ chain and the $XX$ ladder with a superconducting qubit chain and ladder, respectively, on a programmable quantum processor consisting of 24 qubits. Through the measurements of local observables and von Neumann entanglement entropy, we observe two distinct non-equilibrium dynamical behaviors of the qubit chain and ladder. Specifically, during the dynamics of the qubit ladder, the results of local observables validate the predictions of the Gibbs ensemble. Moreover, entanglement entropy saturates the maximum value corresponding to the average entropy of subsystems in random pure states~\cite{EE4}. However, with these signatures of thermalization, the dynamics of the $XX$ chain is verified to be nonergodic due to its integrability. Furthermore, without the need of time-reversal operations for measuring the widely explored OTOCs, by performing efficient and accurate quantum state tomography (QST), we monitor the quench dynamics of the tripartite mutual information (TMI) as a genuine quantification of information scrambling~\cite{scrambling_xiaoliang}. For the first time, we present a critical experimental evidence of scrambling, characterized by a stable negative value of TMI in the $XX$ ladder.


Our experiments are performed on a ladder-type superconducting circuit comprised of 24 transmon qubits (see Fig. \ref{exp_setup}). The superconducting circuit can be described by a Bose-Hubbard Hamiltonian~\cite{BH2}
\begin{eqnarray}
\hat{H} \nonumber &=& \sum_{m\in\{1,2\}}\sum_{n=1}^{11} J_{mn}^{\parallel} (\hat{a}_{m,n}^{+}\hat{a}_{m,n+1}^{-} + \text{h.c.}) \\ \nonumber
&+& \sum_{n=1}^{12} J_{n}^{\perp}(\hat{a}_{1,n}^{+}\hat{a}_{2,n}^{-} + \text{h.c.})+ \sum_{m\in\{1,2\}}\sum_{n=1}^{12}\mu_{mn}\hat{N}_{m,n} \\ 
&+& \sum_{m\in\{1,2\}}\sum_{n=1}^{12}\frac{U_{mn}}{2} \hat{N}_{m,n}(\hat{N}_{m,n} - 1),
\label{SQP}
\end{eqnarray}
with $m$ denoting the number of rung, $\hat{a}_{m,n}$ ($\hat{a}_{m,n}^{\dagger}$) as the bosonic annihilation (creation) operator, $\hat{N}_{m,n} = \hat{a}^{\dagger}_{m,n}\hat{a}_{m,n}$ as the bosonic number operator, $\mu_{mn}$ and $U_{mn}$ denoting the on-site chemical potential and nonlinear interaction, and $J_{n}^{\perp}$ and $J_{mn}^{\parallel}$ referring to the rung and intrachain hopping interactions, respectively.

Since $|\overline{U}|/\overline{J}\simeq 19$ with $|\overline{U}|$ and $\overline{J}$ being the average value of nonlinear and hopping interactions (see Supplementary Information), the system (\ref{SQP}) approximates to the $XX$ spin model where the bosonic annihilation and creation operator are mapped to the spin lowering and raising operator, i.e., $\hat{a}^{\dagger} (\hat{a}) \rightarrow \hat{\sigma}^{+} (\hat{\sigma}^{-})$~\cite{BH3}. Thus, the qubit chain can be described by  $\hat{H}_{\text{C}}= \overline{J}\sum_{n} (\hat{\sigma}_{n}^{+}\hat{\sigma}_{n+1}^{-} + \text{h.c.})$, transformed to a quadratic fermionic model using Jordan-Wigner transformation~\cite{XY_chain}. However, the $XX$ ladder cannot be written as a quadratic form~\cite{XX_ladder2}, which is an interacting fermionic model.

\begin{figure}
	\centering
	\includegraphics[width=1\linewidth]{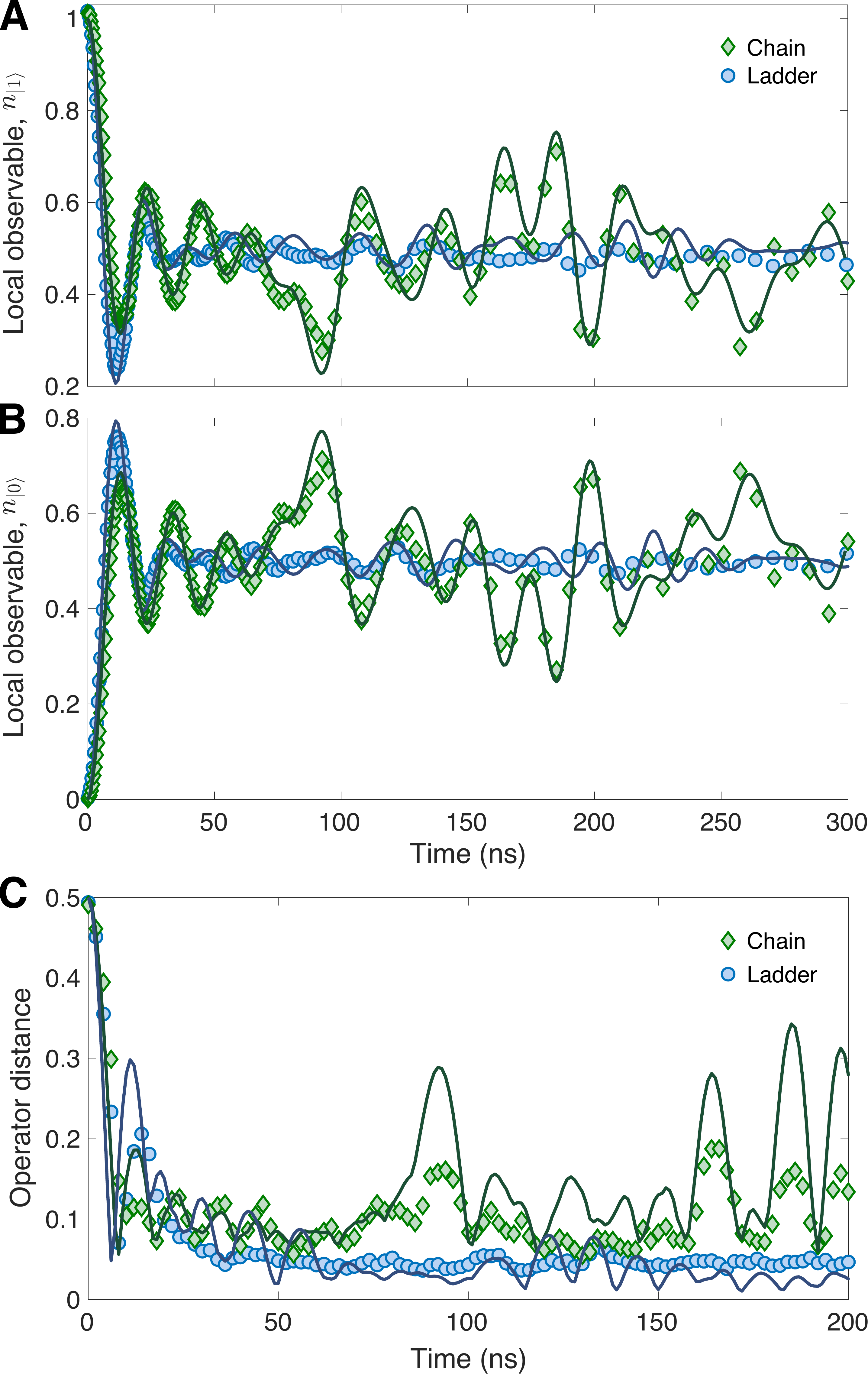}\\
	\caption{\textbf{Dynamics of local densities and the operator distances.} (\textbf{A} and \textbf{B}) Experimental data of the time evolution of local observables $n_{|1(0)\rangle} (t)$ in the qubit chain and ladder. (\textbf{C}) The time evolution of the operator distance between the quenched and thermal states in the chain and ladder. The solid lines are numerics without considering decoherence. }\label{local_obseravble}
\end{figure}

\begin{figure}
	\centering
	\includegraphics[width=1\linewidth]{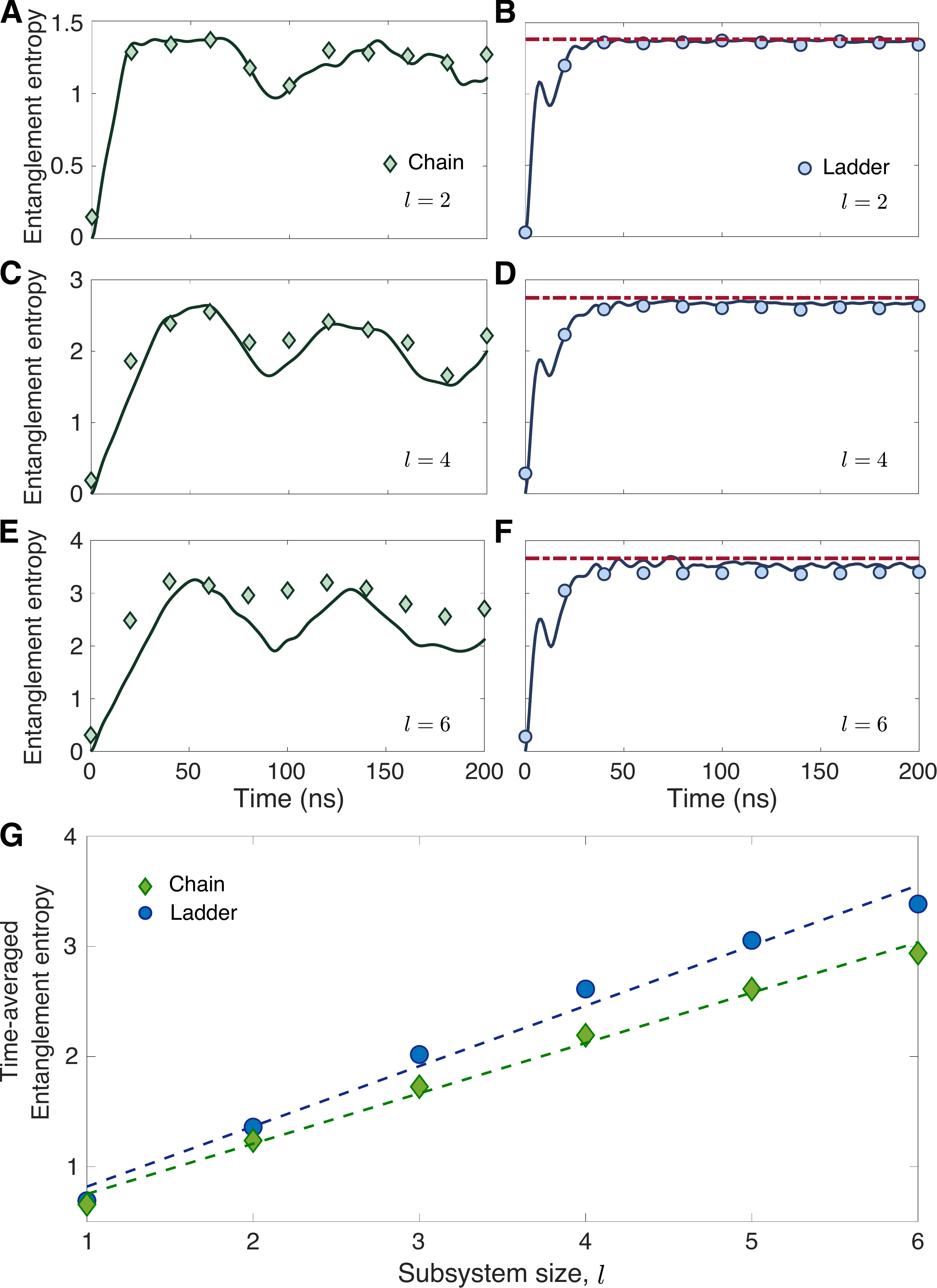}\\
	\caption{\textbf{Dynamics of entanglement entropy with different subsystem length $l$.} (\textbf{A}) Time evolution of EE in the qubit chain with the subsystem consisting of $Q_{1}$ and $Q_{2}$. (\textbf{B}) Time evolution of EE in the qubit ladder with the subsystem consisting of $Q_{1}$ and $Q_{2}$. (\textbf{C}) and (\textbf{D}) are similar to (\textbf{A}) and (\textbf{B}), but with the subsystem consisting of $Q_{1}$--$Q_{4}$. (\textbf{E}) and (\textbf{F}) are similar to (\textbf{A}) and (\textbf{B}), but with the subsystem consisting of $Q_{1}$--$Q_{6}$. (\textbf{G}) The time-averaged EE as a function of  $l$. The solid lines in (\textbf{A} to \textbf{F}) are numerics without considering decoherence. The discrepancy between the experimental and numerical results in (\textbf{E}) can be interpreted as the influence of decoherence (see Supplementary Information). The dashed lines in (\textbf{B}), (\textbf{D}) and (\textbf{F}) denote the Page value of von Neumann EE $S_{\text{Page}} = \ln m - m/2n$ with $m=2^{l}$ and $n=2^{N-l}$ ($N=12$ as the number of qubit). The dashed lines in (\textbf{G}) are the linear fittings of the experimental data. }\label{ee}
\end{figure}

\begin{figure*}
	\centering
	\includegraphics[width=1\linewidth]{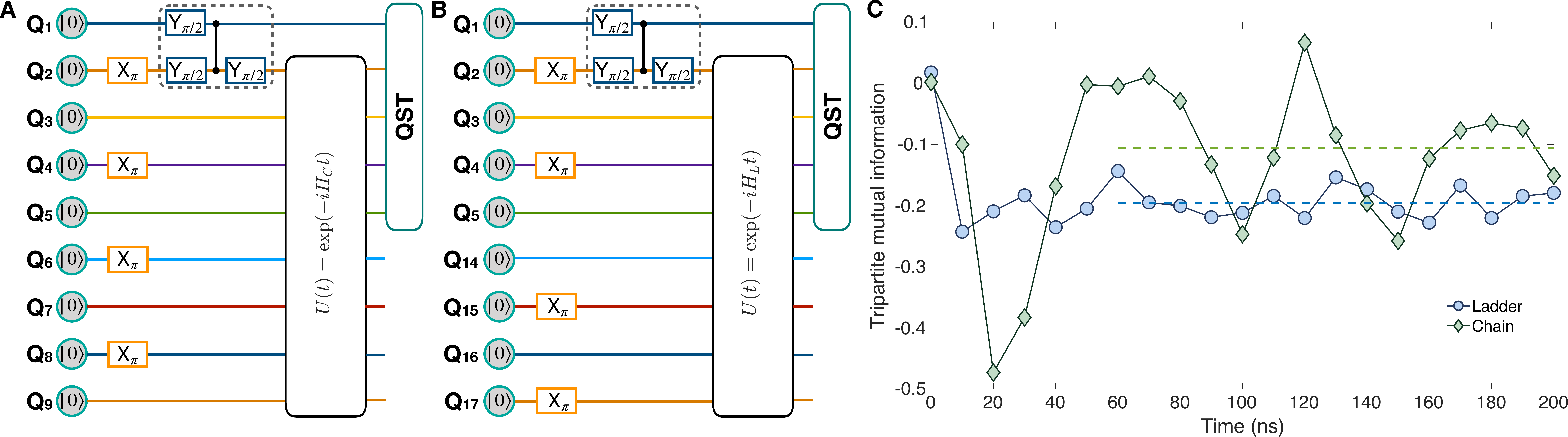}\\
	\caption{\textbf{Information scrambling quantified by the tripartite mutual information.} (\textbf{A}) The schematic experimental pulse sequence for the dynamics of tripartite mutual information in the qubit chain. (\textbf{B}) is similar to (\textbf{A}) but in the qubit ladder. (\textbf{C}) The experimental data of the time evolution of tripartite mutual information in the qubit chain and ladder. For the qubit chain and ladder, the averaged values are $-0.106$ and $-0.196$ (highlighted by the dashed lines) respectively. The comparison with numerics for the experimental data is presented in the Supplementary Information.}\label{tmi}
\end{figure*}

To probe ergodic dynamics, we consider the local observable $\hat{n}_{|1(0)\rangle}=(\sum_{m\in N_{1(0)}} \hat{\sigma}_{m}^{+}\hat{\sigma}_{m}^{-})/N_{1(0)}$, summing over the $N_{1(0)}$ qubits initialized in $|1\rangle$ ($|0\rangle$) and averaging it. Applying the pulse sequence in Fig.~\ref{exp_setup}C and D, we can monitor the dynamics of local observables (known as local densities~\cite{BH3}) via 3,000 repeated single-shot measurements.
If the dynamics is ergodic, local densities will approach to a stationary value $1/2$ after a short relaxation. In Fig. \ref{local_obseravble}A and B, it is shown that local densities converge to $1/2$ after $t\simeq 30$ ns in the ladder, which is a signature of thermalization. Whereas, the convergence cannot be observed in the chain until $t\simeq 300$ ns. This experimental data of local densities in the chain are consistent with the analytical results of the 1D Bose-Hubbard model with the limit case of the nonlinear interaction $U=\infty$.~\cite{BH3} (see Supplementary Information).

We then study ergodicity via the operator distance $d(\rho_{A}(t),\rho_{A}^{(\text{eq})}(T))$ as the maximum eigenvalue of $\rho_{A}(t) - \rho_{A}^{(\text{eq})}(T)$, where $\rho_{A}(t)$ is the single-site reduced density matrix at time $t$ measured using the QST, and $\rho_{A}^{(\text{eq})}(T)$ is the Boltzmann density operator with temperature $T\rightarrow\infty$ (see Supplementary Information). When the dynamics is ergodic, it can be predicted that $d(\rho_{A}(t),\rho_{A}^{(\text{eq})}(T))=0$ for a long time $t$~\cite{thermalization_mbl,cold_atom}. Figure~\ref{local_obseravble}C shows the time evolutions of the $d(\rho_{A}(t),\rho_{A}^{(\text{eq})}(T))$ averaged over all qubits. The distance shows a value smaller than 0.05 for the ladder, while it exhibits a strong oscillation between 0.1 and 0.2 for the chain, providing an evidence of the occurrence and absence of ergodicity in the $XX$ ladder and chain, respectively. 

We also investigate the entanglement entropy (EE), as a quantification of bipartite entanglement, characterizing ergodicity via the volume law extracted from its dependence on the subsystem size $l$~\cite{EE4,EE5}. Indirect methods of measuring the second R\'{e}nyi EE, including quantum interference~\cite{cold_atom} and randomized measurements~\cite{EE_exp1}, have been developed. Nonetheless, the measurement of the von Neumann EE requires the accurate and efficient QST. We perform a 6-qubit state tomography to obtain the reduced density matrix $\rho_{A}(t)$ with the subsystem $A$ comprised of $Q_{1}$--$Q_{6}$, and then calculate the EE $S_{A}=-\text{Tr}[\rho_{A}(t)\ln \rho_{A}(t)]$. By partially tracing the 6-qubit density matrix, we also obtain the EE of smaller subsystems.

Figure~\ref{ee}A--F shows the dynamics of the EE in the qubit chain and ladder. We observe that the temporal fluctuations of EE become more dramatic in the chain than that in the ladder. Furthermore, we study the time-averaged EE (after $t=60$ ns) as a function of the subsystem size $l$. As depicted in Fig.~\ref{ee}G, the volume law of EE $S_{A}\propto l$ is satisfied for the quenched states in both qubit chain and ladder. However, the value of EE is larger for the ladder, which approaches to the Page value for random pure states~\cite{EE4}. In short, the experimental data of EE are consistent with the results in Ref.~\cite{EE5}, where stronger fluctuations and a smaller volume-law slope in integrable systems than those in non-integrable cases are revealed.

Next, we study information scrambling by considering tripartite mutual information (TMI)~\cite{scrambling_xiaoliang}:
\begin{eqnarray}
I_{3} &= & S(\rho_{A}) + S(\rho_{B}) + S(\rho_{C}) + S(\rho_{ABC}) \\ \nonumber
&-& S(\rho_{AB}) - S(\rho_{AC}) - S(\rho_{BC}),
\label{tmi_def}
\end{eqnarray}
where $S(\rho)$ is the von Neumann entropy, and $A$, $B$ and $C$ refer to three subsystems. Experimentally, to calculate TMI, we measure $\rho_{ABC}$ using QST, and obtain the density matrix of smaller subsystems by partially tracing $\rho_{ABC}$.

The schematic experimental pulse sequence for measuring TMI in the qubit chain is depicted in Fig.~\ref{tmi}A.
Different from the previous pulse sequences (Fig.~\ref{exp_setup}C and D), the qubits $Q_{1}$ and $Q_{2}$ are prepared in a Einstein-Podolsky-Rosen (EPR) pair $|\text{EPR}\rangle_{12} =\frac{1}{\sqrt{2}}(|0\rangle_{1}|0\rangle_{2} +|1\rangle_{1}|1\rangle_{2} ) $ by the $\text{X}_{\pi}$ and a CNOT gates (see the frames in Fig.~\ref{tmi}A and B). Subsystems $A$ and $B$ are chosen as $Q_{1}$ and $Q_{2}$, respectively, and the subsystem $C$ consists of $Q_{3}$--$Q_{5}$. A similar scheme of scrambling in the qubit ladder is plotted in Fig.~\ref{tmi}B with the same choice of subsystem $A$, $B$ and $C$, but the remainder becomes $Q_{14}$--$Q_{17}$. The initialization protocol in Fig.~\ref{tmi}A and B are enlightened by the quantum teleportation and information retrieval from black holes~\cite{scrambling_pra,tele_1,tele_2}, and the dynamics of TMI can characterize how the local information encoded by the EPR pair scrambles.

Figure~\ref{tmi}C shows the experimental results of the quench dynamics of TMI for the qubit chain and ladder. In the qubit chain as an integrable case, TMI recovers zero after the decreasing period, while in the qubit ladder, TMI saturates to a stationary negative value. Moreover, for the $XX$ ladder, the value of time-averaged TMI (after $t=60$ ns), smaller than that in the chain, reflects a stronger information scrambling.

The measurement of TMI characterizing information scrambling lays the foundation for further experimental studies on TMI in other systems such as digital quantum circuits simulating black holes~\cite{tele_2}. The ladder-type superconducting processor, where ergodicity is observed, can be a suitable platform for experimentally probing the phenomena of ergodicity breaking, such as many-body localization~\cite{thermalization_mbl}, measurement-induced disentangling phase~\cite{measurement_QPT2}, and quantum many-body scars~\cite{scars}.

\begin{acknowledgments}
\noindent The authors thank the USTC Center for Micro- and Nanoscale Research and Fabrication. The authors also thank QuantumCTek Co., Ltd. for supporting the fabrication and the maintenance of room temperature electronics. This research was supported by the National Key R$\&$D Program of China (Grants No. 2018YFA0306703, No. 2017YFA0304300, No. 2016YFA0302104, No. 2016YFA0300600), the Chinese Academy of Sciences, and Shanghai Municipal Science and Technology Major Project (Grant No. 2019SHZDZX01), the Strategic Priority Research Program of Chinese Academy of Sciences (Grant No. XDB28000000), Japan Society for the Promotion of Science (JSPS) Postdoctoral Fellowship (Grant No. P19326), JSPS KAKENHI (Grant No. JP19F19326), the National Natural Science Foundation of China (Grants No. 11574380, No. 11905217, No. 11934018, No. 11774406), Key-Area Research and Development Program of Guangdong Province (Grant No. 2020B0303030001), and Anhui Initiative in Quantum Information Technologies. 

\end{acknowledgments}

\noindent \textbf{Competing interests:} The authors declare no competing interests. 

\noindent \textbf{Data availability:} All relevant data are available from the corresponding authors upon request. 

\noindent \textbf{Author contributions:} H.F., X.Z., and J.-W.P. conceived the research. Q.Z., Z.-H.S., M.G. and X.Z. designed the experiment. Q.Z. designed the sample. Q.Z., H.D., and H.R. prepared the sample. Q.Z. and M.G. carried out the measurements. Y.W. developed the programming platform for measurements. Z.-H.S. and C.Z. did numerical simulations. Q.Z., Z.-H.S., M.G., F.C., Y.-R.Z. and Y.Y. analyzed the results. Q.Z., Z.-H.S., M.G., Y.-R.Z., H.F., X.Z. co-wrote the manuscript. J.L., Y.X., L.S., C.G., F.L., and C.-Z.P. developed room temperature electronics equipments. All authors contributed to discussions of the results and development of manuscript. X.Z. and J.-W.P. supervised the whole project. 

%

\clearpage
\renewcommand\thefigure{S\arabic{figure}}
\renewcommand\thetable{S\arabic{table}}

\noindent \emph{\textbf{Supplementary Materials for `Observation of thermalization and information scrambling in a superconducting quantum processor'}}

\section{Device}
\subsection{Architecture}

\begin{figure*}[!htb]\centering
	\includegraphics[width=0.9\linewidth]{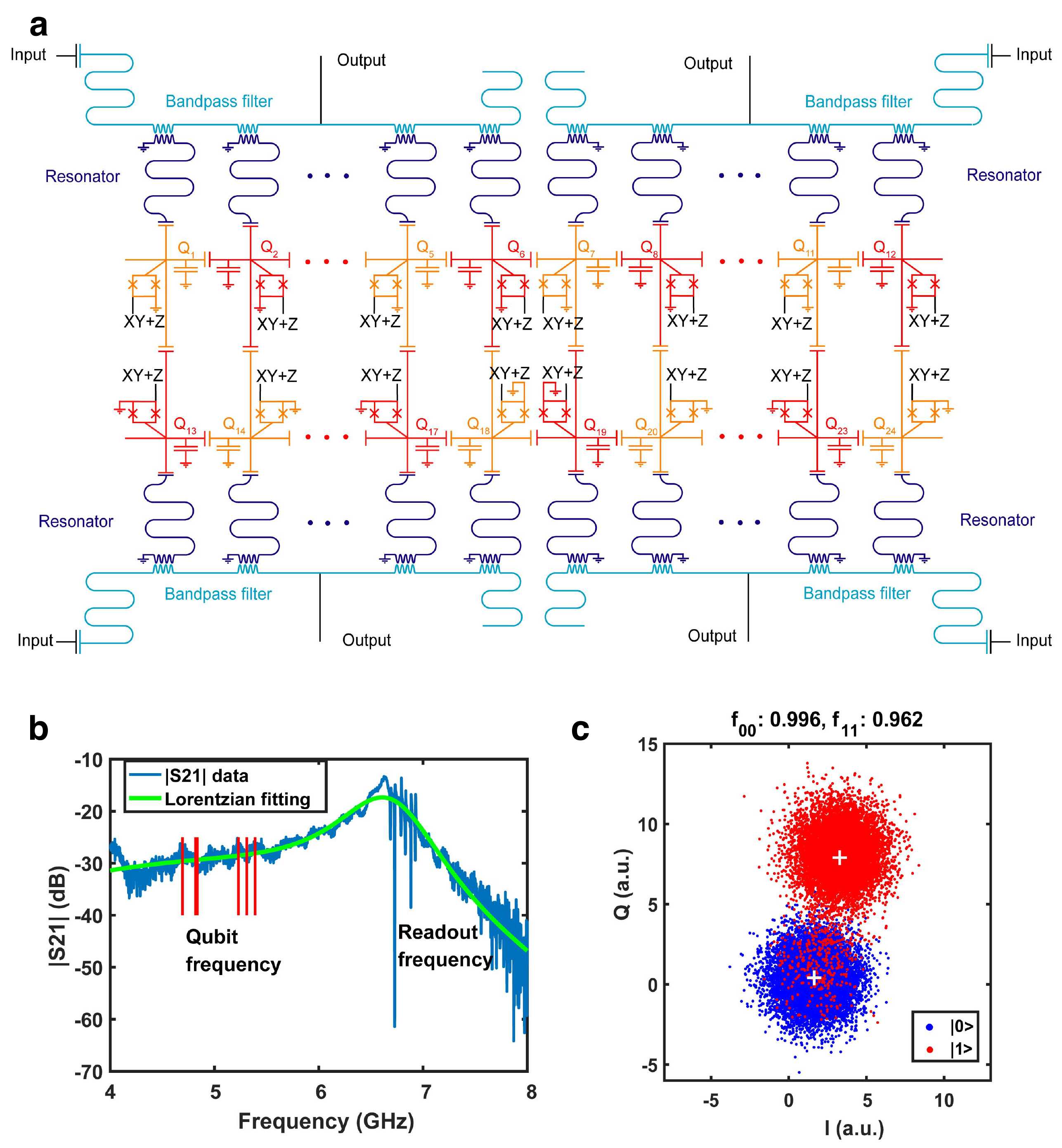}
	\caption{\textbf{Superconducting quantum circuit and the state readout.} \textbf{a} Circuit diagram of the device. Each qubit is capacitively coupled to its nearest neighbors, inductively coupled to a control line for XY and Z control, and dispersively coupled to a $\lambda/4$ resonator for readout. Specially, every six readout resonators are inductively coupled to a $\lambda/2 $ bandpass filter for fast and high fidelity multi-qubit readout. Each bandpass filter has a capacitively coupled input line and a inductively coupled output line. \textbf{b} Transmission spectrum of the bandpass filter for the readout resonators of $Q_{1}-Q_{6}$, measured with a vector network analyzer. The $\lambda/2 $ bandpass filter $|S21|$ data has a shape of Lorentzian type, which is designed to enhance the readout frequencies and suppress the qubit frequencies. \textbf{c} Single shot readout events for $Q_{4}$. The readout process is repeated for 10,000 times after the qubit is successfully initialized in $\ket{0}$ (blue dots) and $\ket{1}$ (red dots), respectively. The fidelity $f_{00}$ is $99.6$ $\textrm{\%}$, and $f_{11}$ is $96.2$ $\textrm{\%}$. }  \label{exp_supp_1}
\end{figure*}

Our device is a 24-qubit superconducting quantum processor arranged into two rows of $12$ Transmon qubits~\cite{BH2}. The simplified circuit diagram of the device is shown in Fig.~\ref{exp_supp_1}\textbf{a}. We use $18$ of them for this experiment. The qubits $Q_1$ - $Q_{12}$ are used for the quantum simulation of the 1D $XX$ chain, and the qubits $Q_1$ - $Q_6$ and $Q_{13}$ - $Q_{18}$ are used for the quantum simulation of the $XX$ ladder. With the hard-core boson limit~\cite{BH3}, the Hamiltonian of the qubit chain can be written as 
\begin{eqnarray}
\hat{H}_{\text{C}} = \sum_{n=1}^{11} J^{\parallel}_{n}(\hat{\sigma}_{n}^{+}\hat{\sigma}_{n+1}^{-} + \text{h.c.})
\label{HC}
\end{eqnarray}
with $J^{\parallel}_{n}$ as the intrachain hopping interaction. Similarly, the Hamiltonian of the qubit ladder reads
\begin{eqnarray}\nonumber
\hat{H}_{\text{L}} &=& \sum_{m\in\{1,2\}}\sum_{n=1}^{5} J^{\parallel}_{mn} (\hat{\sigma}_{m,n}^{+}\hat{\sigma}_{m,n+1}^{-} + \text{h.c.}) \\ 
&+& \sum_{n=1}^{6} J_{n}^{\perp} (\hat{\sigma}_{1,n}^{+}\hat{\sigma}_{2,n}^{-} + \text{h.c.}),
\label{HL}
\end{eqnarray}
where $m$ is the number of rung, and $J_{n}^{\perp}$ and $J^{\parallel}_{mn}$ refer to the rung and intrachain interactions. 

Each qubit is capacitively coupled to its nearest neighbors with a fixed coupling strength. The qubit energy relaxation time $T_1$, and dephasing time $T_2^*$ are presented in Table~\ref{tab1}, which are measured at the idle frequency $\omega_{\textrm{q}}^{\textrm{idle}}/2\pi$. The coupling strength $J/2\pi$ measured at the working frequency $4.863$~GHz by two-qubit resonant oscillations are presented in Table~\ref{tab1}, which is about $12.3$~MHz for intrachain interactions ($J_{n}^{\parallel}$) and $13.6$~MHz for rung interactions ($J_{n}^{\perp}$). We define the rate of correctly measuring $|1\rangle$ ($|0\rangle$) when the qubit is prepared at $|1\rangle$ ($|0\rangle$) as $f_{11}$ ($f_{00} $). As an example, the single shot events of $Q_4$ is shown in Fig.~\ref{exp_supp_1}\textbf{c}. After the integration time of $1100$ ns, the fidelity of $f_{00}$ and $f_{11}$ is determined as $99.6$ $\textrm{\%}$, and $96.2$ $\textrm{\%}$, respectively.

\subsection{Readout and bandpass filter}
For state readout we dispersively couple each qubit to a $\lambda/4$ readout resonator with coupling strengths $g_{\textrm{res}}/2\pi$ designed to be about $100$~MHz. The measured resonator frequencies $\omega_{\textrm{read}}/2\pi$ and coupling strengths $g_{\textrm{res}}/2\pi$ are listed in Table~\ref{tab1}. Resulting from the increasing of the coupling strength, the Purcell effect formed by the readout line is non-negligible~\cite{Purcell1946}. To mitigate that side effect, we insert a $\lambda/2 $ bandpass filter~\cite{Reed2010,Jeffrey2014,Sete2015} between the readout line and the resonators. The bandpass filter suppresses the coupling of qubit frequencies while enhances the coupling of readout frequencies, as show in Fig.~\ref{exp_supp_1}\textbf{b}. As a result, the bandpass filter allows for fast and high fidelity readout while maintaining the high-quality qubit performance by reducing the environmental damping. The bandpass filter is designed with a bandwidth of about $300$~MHz, which covers the spanning of six readout resonators. The leakage time $1/{\kappa_{r}}$ of the readout resonator after coupled with the bandpass filter is designed to be about $100$~ns. However, as a result of the frequency drift in fabrication, those resonators whose frequencies are far away from the center frequency of the bandpass filter have larger values of $1/{\kappa_{r}}$. More detailed parameters can be found in Table~\ref{tab1}.

\begin{table*}[t]
	\centering
	\scriptsize
	\renewcommand\arraystretch{1.8}
	\resizebox{\textwidth}{!}{
		\begin{tabular}{c |c c c c c c c c c c c c c c c c c c c}
			\hline
			\hline
			
			&$\text{Q}_{1}$ &$\text{Q}_{2}$ &$\text{Q}_{3}$ &$\text{Q}_{4}$ &$\text{Q}_{5}$ &$\text{Q}_{6}$			
			&$\text{Q}_{7}$ &$\text{Q}_{8}$ &$\text{Q}_{9}$ &$\text{Q}_{10}$ &$\text{Q}_{11}$ &$\text{Q}_{12}$
			&$\text{Q}_{13}$ &$\text{Q}_{14}$ &$\text{Q}_{15}$ &$\text{Q}_{16}$ &$\text{Q}_{17}$ &$\text{Q}_{18}$\\
			\hline
			$\omega_{\textrm{read}}/2\pi$~(GHz)
			&6.688	&6.729 &6.790	&6.832	&6.885	&6.927	&6.701	&6.755	&6.801	&6.855	&6.908	&6.954 &6.384	&6.423	&6.482	&6.521	&6.576	&6.637\\
			$\omega_{\textrm{q}}^{\textrm{max}}/2\pi$~(GHz)
			&4.928  &5.536	&4.962  &5.600  &4.887	&5.600  &4.941	&5.562	&4.904	&5.602	&4.905  &5.587 &5.562  &4.89	&5.571  &4.902  &5.525	&4.928\\
			$\omega_{\textrm{q}}^{\textrm{idle}}/2\pi $~(GHz)
			&4.835  &5.31	&4.693	&5.39	&4.82	&5.23	&4.68	&5.32	&4.77	&5.25	&4.67	&5.42 &5.377  &4.74	&5.47	&4.88	&5.29	&4.76\\
			$T_1$~($\mu$s)
			&24.3  	&22.8	&26.5	&24.0	&28.8	&25.9	&19.5	&28.5	&20.5	&17.9	&31.8	&13.1 &16.6  	&22.4	&12.4	&24.4	&23.9	&21.4\\
			$T_2^*$~($\mu$s)
			&5.2    &2.0	&1.8    &2.2	&6.2    &1.8    &5.5    &2.3    &4.1	&2.0    &10.4	&2.3  &2.7    &2.9	&4.6    &10.4	&2.5    &3.2\\		
			$U/2\pi $~(MHz)
			&$-238$ &$-230$ &$-240$ &$-230$ &$-238$  &$-230$ &$-240$ &$-230$ &$-236$ &$-230$&$-236$ &$-230$ &$-226$ &$-236$	&$-226$ &$-236$ &$-228$&$-236$ \\
			$g_{\textrm{res}}/2\pi$~(MHz)
			&111	&112	&115	&112	&110	&119	&109	&110	&108	&113	&116	&116  & 103   &105   &106   &107    &106     &110\\
			$1/{\kappa_{r}}$~(ns)
			&65  	&63 	&92 	&65 	&99 	&125	&67 	&62 	&69 	&111	&172	&263   &62   &63     &75 	&64     &83  	&149\\
			$n_{\textrm{photons}}$
			&19     &11 	&25     &11 	&32    &38     &24      &13     &15	     &22    &47	    &156   &13    &13	 &5     &22	     &11     &62\\
			$f_{00}$~($\textrm{\%}$)
			&99.0   &99.8	&99.4   &99.6	&99.0  &99.5   &99.7    &99.9	&99.7    &99.7	&98.9   &98.1  &99.7   &99.7  &99.8  &99.5   &97.5	&97.5\\
			$f_{11}$~($\textrm{\%}$)
			&89.0   &94.2	&95.0   &96.2	&91.7  &93.9   &94.2   &93.7    &91.4	&96.7    &95.5	&93.0  &90.7   &90.9  &94.7  &93.8	 &92.4   &94.1\\
			$\textrm{Integration time}$ (ns)
			&1700   &1100	&1200   &1100	&1400  &1500   &1400   &1250    &1200	&1000    &1300	&1500  &1100   &1200  &900  &900	 &1000   &1000\\
			\hline
			$1$Q XEB fidelity~($\textrm{\%}$)
			&99.91 &99.87 &99.92 &99.86 &99.91 &99.88 &99.87 &99.88 &99.84 &99.75 &99.88 &99.87 &99.89 &99.70 &99.72 &99.85 &99.85 &99.94\\
			$1$Q SPB fidelity~($\textrm{\%}$)
			&99.92 &99.89 &99.92  &99.89 &99.92 &99.87 &99.88 &99.89 &99.84	&99.76 &99.91 &99.88 &99.89 &99.74 &99.74 &99.90 &99.90 &99.94\\
			\hline
			
			&\multicolumn{18}{c}{~~$\text{Q}_{1}$$\text{Q}_{2}$ ~~~~$\text{Q}_{2}$$\text{Q}_{3}$ ~~~~$\text{Q}_{3}$$\text{Q}_{4}$
				~~~~$\text{Q}_{4}$$\text{Q}_{5}$ ~~~~$\text{Q}_{5}$$\text{Q}_{6}$ ~~~~$\text{Q}_{6}$$\text{Q}_{7}$
				~~~~$\text{Q}_{7}$$\text{Q}_{8}$ ~~~~$\text{Q}_{8}$$\text{Q}_{9}$ ~~~~$\text{Q}_{9}$$\text{Q}_{10}$
				~$\text{Q}_{10}$$\text{Q}_{11}$ ~$\text{Q}_{11}$$\text{Q}_{12}$} \\
			$J/2\pi $~(MHz)
			&\multicolumn{18}{c}{~~~~~~12.2~~~~~~~~~12.2~~~~~~~~12.1~~~~~~~~12.0~~~~~~~~12.3~~~~~~~~13.2~~~~~~~~12.5~~~~~~~~12.5~~~~~~~~12.2~~~~~~~~12.2~~~~~~~~12.2~~~~~~} \\
			\hline
			
			&\multicolumn{18}{c}{$\text{Q}_{13}$$\text{Q}_{14}$~$\text{Q}_{14}$$\text{Q}_{15}$ ~$\text{Q}_{15}$$\text{Q}_{16}$
				~$\text{Q}_{16}$$\text{Q}_{17}$ ~$\text{Q}_{17}$$\text{Q}_{18}$ ~~$\text{Q}_{1}$$\text{Q}_{13}$
				~~$\text{Q}_{2}$$\text{Q}_{14}$ ~~$\text{Q}_{3}$$\text{Q}_{15}$ ~~$\text{Q}_{4}$$\text{Q}_{16}$
				~~$\text{Q}_{5}$$\text{Q}_{17}$ ~~$\text{Q}_{6}$$\text{Q}_{18}$}\\
			$J/2\pi $~(MHz)
			&\multicolumn{18}{c}{~~~~~~12.4~~~~~~~~~12.3~~~~~~~~12.4~~~~~~~~12.4~~~~~~~~12.2~~~~~~~~13.3~~~~~~~~13.6~~~~~~~~13.7~~~~~~~~13.8~~~~~~~~13.7~~~~~~~~13.6~~~~~~} \\
			\hline
			\hline
		\end{tabular}	
	}
	
	\caption{
		{\textbf{Parameters of the device:} $\omega_{\text{read}}/2\pi$ is the frequency of the readout resonator; $\omega_{\text{q}}^{\text{max}}/2\pi$ is the maximum frequency of the qubit; $\omega_{\text{q}}^{\text{idle}}/2\pi$ is the idle frequency of the qubit; $T_1$ and $T_2^*$ are the energy relaxation and dephasing time of the qubit, respectively, which are measured at the idle frequency; $U$ is the anharmonicity of qubit measured at the idle frequency; $1/{\kappa_{r}}$ is the leakage time of the readout resonator; $J/2\pi $ is the coupling strength of the corresponding qubit-pair measured at the working frequency ($4.863$~GHz); $n_{\text{photons}}$ represents the number of photons occupied in the resonator calibrated by measuring the ac Stark shift of the qubit; $f_{11}$ ($f_{00}$) is the probability of correctly identifying the qubit state when it is initially prepared in $|1\rangle$ ($|0\rangle$); The $1$Q XEB fidelity is the average gate fidelity of single-qubit $\text{X}/2$ gate measured at idle frequency. The $1$Q SPB fidelity describes the effect of decoherence on the single-qubit $\text{X}/2$ gate. The length of $\text{X}/2$ gates for all qubits is $30$~ns. }
	}
	\label{tab1}	
\end{table*}

\subsection{Fabrication of airbridge}

We use HF airbridges~\cite{Dunsworth2018}, instead of crossovers, to connect the control lines separated by bandpass filters.
In addition, the HF airbridges are used across the control lines, readout resonators and bandpass filters to suppress parasitic slotline modes. The device is fabricated in the same way as previous sample~\cite{ladder_sqp_ye,Yan2019}, except for the process in the fabrication of HF airbridges.

\begin{figure*}[!htb]\centering
	\includegraphics[width=0.9\linewidth]{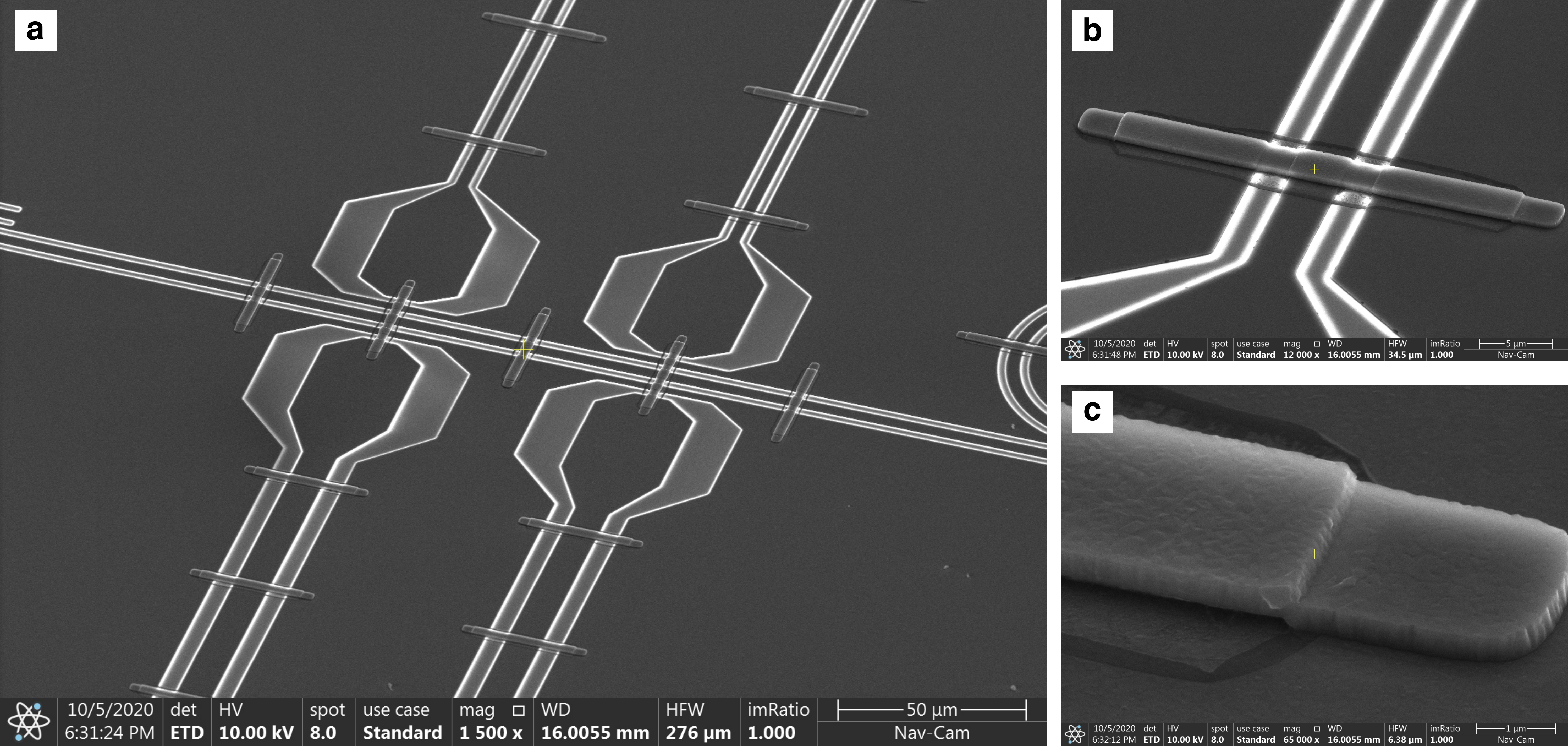}
	\caption{\textbf{SEM photographs of HF airbridges.} \textbf{a} HF airbridges are used between those control lines separated by bandpass filters to connect them, and across the control lines, readout resonators and bandpass filter to suppress parasitic slotline modes. \textbf{b} An HF airbridge across the control line. \textbf{c} One end of the HF airbridge in \textbf{b}. }  \label{exp_supp_2}
\end{figure*}

Here we briefly describe the fabrication of HF airbridges. First, a $500$~nm SiO$_2$ dielectric layer is defined by laser lithography followed by electron-beam evaporation. Second, the upper $500$~nm aluminum electrodes are fabricated with laser lithography and electron-beam evaporation. Lastly, HF airbridges are fabricated with a dry VHF etcher to remove the dielectric layer. In short, the fabrication of HF airbridges is the same as that of crossovers, except for the final step which removes the dielectric layer. A scanning electron micrograph (SEM) photograph of HF airbriges is shown in Fig.~\ref{exp_supp_2}.

\section{Experimental wiring setup}

\begin{figure}[!htb]\centering
	\includegraphics[width=1\linewidth]{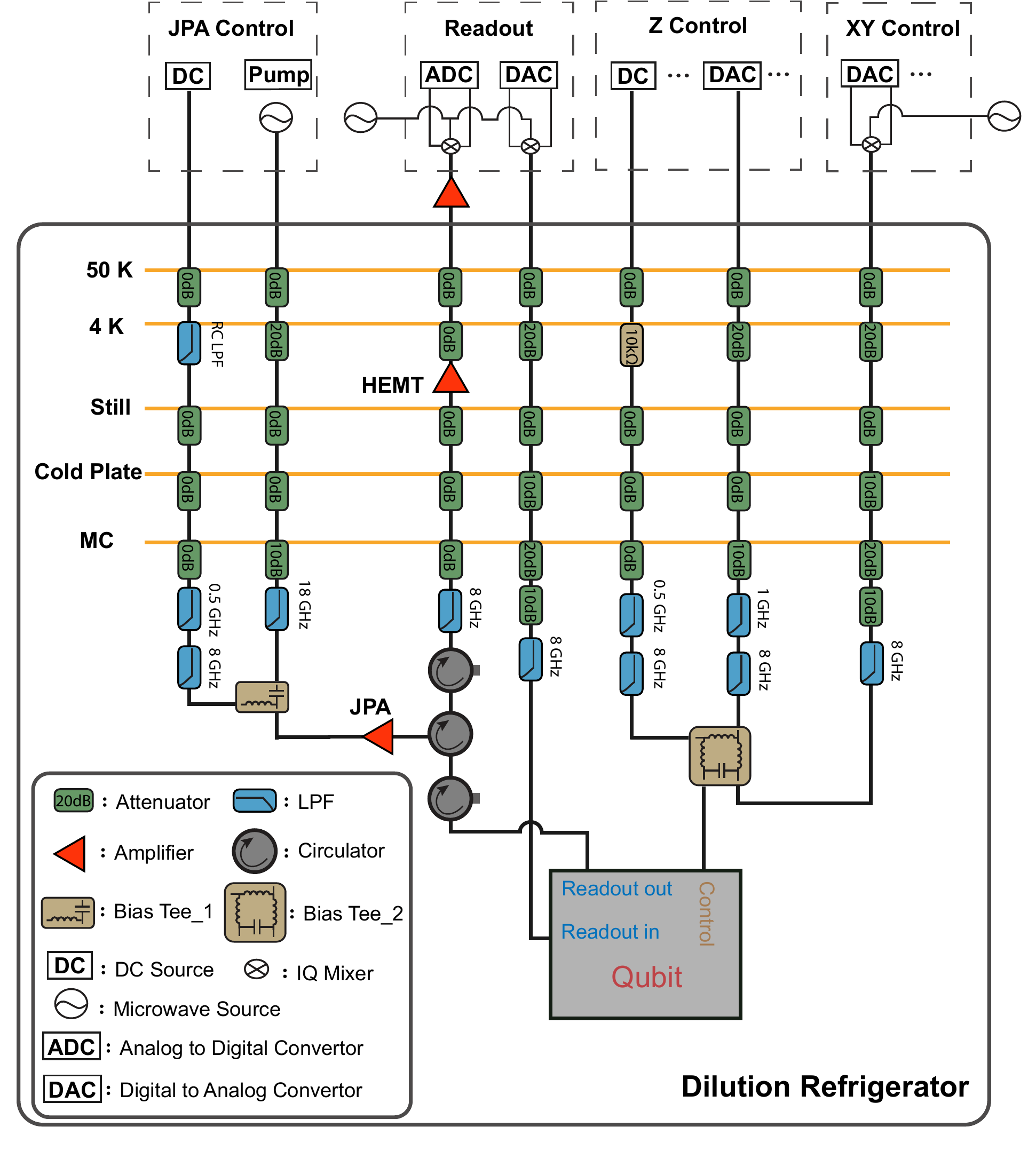}
	\caption{\textbf{Schematic of room temperature electronics and cryogenic wiring setup.} }  \label{exp_supp_3}
\end{figure}

The experimental wiring setup for qubit control and frequency-multiplexed readout at different stages of the cryogenic system and the schematic of room temperature electronics are shown in Fig.~\ref{exp_supp_3}. The quantum processor device is installed under the mixing chamber of the dilution refrigerator (DR), whose base temperature is about $12$~mK, with a magnetic field shield. The readout and control waveforms are generated by Digital-to-Analog Converters (DAC) at room temperature, and then attenuated by different attenuators installed at different stages of the DR. The signals are finally filtered by different low-pass filters installed under the mixing chamber plate. The DC signals are damped by $10$~K$\Omega$ resistors installed at 4K plate. Before arriving at the control lines of the quantum device, XY, Z and DC controls are combined together by bias-tees. To obtain higher signal-noise ratio (SNR) in state readout, we use Josephson parametric amplifiers (JPA)~\cite{Mutus2014} as the first stage amplification of readout signals. The high-electron-mobility transistors (HEMT) and low-noise microwave amplifiers, working at $4$~K stage and room temperature, are used as the second and third stage amplifications, respectively. The average gain of the JPAs is $8.8$~dB, $14.1$~dB, and $13.2$~dB, respectively. The signal carrying qubits information are finally demodulated and digitized by Analog-to-Digital Converters (ADC). 

The room temperature electronics used in this experiment includes $80$ DAC channels, 6 ADC channels, $28$ DC channels and 7 microwave source channels. Among them, $6$ DC channels are employed to keep the unused qubits ($Q_{19}$, $Q_{20}$, $Q_{21}$, $Q_{22}$, $Q_{23}$, $Q_{24}$) idling  below $4.2$~GHz.

\section{Gate performance}
\subsection{Single-qubit gate}

We use the cross entropy benchmarking (XEB) to benchmark the fidelity of single-qubit $\text{X}/2$ gate~\cite{DgateXEB2019,XEB2018,Google-supermacy2019}. In the single-qubit gate benchmarking process, many cycles of random single-qubit gates are applied. Each cycle consists of one single instance of gate sequence sampled from $80$ random circuits. The circuits use a single-qubit gate set formed by the $\pi/2$ rotations around the eight axes in the Bloch representation: $\pm$$X$, $\pm$$Y$, and $\pm$$(X\pm$$Y)$ , and end with a random single-qubit gate before measurement.

\begin{figure*}[!htb]\centering
	\includegraphics[width=0.8\linewidth]{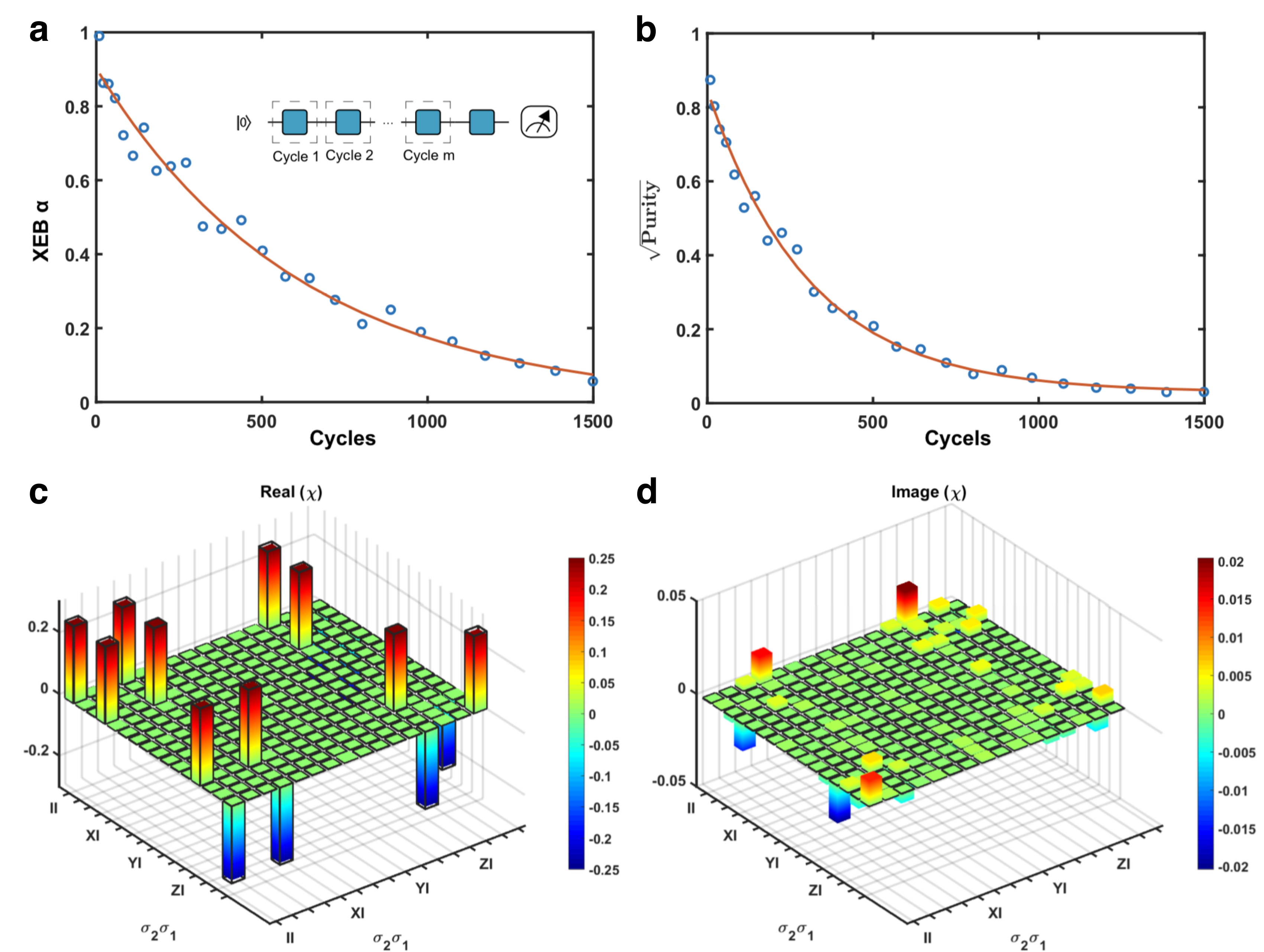}
	\caption{\textbf{Sanity check of gate performances.} \textbf{a} Single-qubit XEB fidelity versus cycles number using $80$ random circuits for $Q_3$. The cycle fidelity is 99.84$\textrm{\%}$, and the average fidelity of single-qubit $\text{X}/2$ gate is 99.92$\textrm{\%}$. Inset: The blue square represents the random single-qubit gates, each cycle is comprised of single-qubit gate sequences from $80$ random circuits, and ended with a random single-qubit gate and measurement. \textbf{b} Single-qubit SPB fidelity versus cycles number extracted from XEB data. The cycle fidelity is 99.68$\textrm{\%}$, and the average SPB fidelity of single-qubit $\text{X}/2$ gate is 99.92$\textrm{\%}$. The result indicates single-qubit $\text{X}/2$ gate error comes mostly from decoherence error. \textbf{c} The real part of the $\chi$ matrix of two-qubit CZ gate between $Q_1$ and $Q_2$. \textbf{d} is similar to \textbf{c} but for the imaginary part. After CPTP projection, the fidelity is {98.7}$\textrm{\%}$, which involves both preparation and measurement errors. }  \label{exp_supp_4}
\end{figure*}

We apply a linear XEB~\cite{Google-supermacy2019} to compare the measured state probabilities with the ideal probabilities, and then acquire the sequence fidelity $\alpha$ as
\begin{eqnarray}
\alpha = \frac{\sum\nolimits_{q}p_m(q)\overline{Dp_s(q)-1}}{D\sum\nolimits_{q}\overline{p_{s}(q)^2}-1},
\label{XEBalpha}
\end{eqnarray}
with $D=2^N(N=1$ for single-qubit XEB), $q$ as one of bitstrings (for single-qubit XEB, $q$ is 0 or 1), $p_s(q)$ as the ideal probability of $q$, and $p_m(q)$ as the measured probability of $q$. The over lines in Eq. (\ref{XEBalpha}) refer to the average of the $80$ random circuits in each cycle.
The sequence fidelity $\alpha$ decaying with the number of cycles $m$ is shown in Fig.~\ref{exp_supp_4}\textbf{a}. The fitting function is $\alpha=Ap^m+B$, where A and B represent the state preparation and measurement errors, respectively. The average error $r$ of single-qubit gate is obtained according to
\begin{eqnarray}
r = (1-p)\frac{D-1}{D},
\label{XEBerror}
\end{eqnarray}
and the average XEB fidelity of single-qubit gate is $F=1-r$.

Meanwhile, we use the speckle purity benchmarking (SPB)~\cite{Google-supermacy2019} to calibrate the effect of decoherence error,
\begin{eqnarray}
\text{Purity} = \text{Var}(P_m)\frac{D^2(D+1)}{D-1},
\label{SPBerror}
\end{eqnarray}
where $\text{Var}(P_m)$ is the variance of the experimental probabilities extracted from the XEB experiment. The fitting function of $\text{Purity}$ versus cycle number $m$ is the same as that of XEB. As an example, the XEB and SPB results of the single-qubit $\text{X}/2$ gate on $Q_3$ are presented in Fig.~\ref{exp_supp_4}\textbf{a} and \textbf{b}, respectively.

\subsection{Two-qubit gate}

In the experiment of probing information scrambling, the qubit $Q_1$ and $Q_2$ are prepared in a Einstein-Podolsky-Rosen (EPR) pair state. The CNOT gate in realizing the entanglement state is realized by a two-qubit controlled-phase (CZ) gate and one single-qubit $\text{Y}_{\pi/2}$ gate on control qubit $Q_1$ and two single-qubit $\text{Y}_{\pi/2}$ gates on target qubit $Q_2$. The two-qubit CZ gate is implemented by tuning the $\ket{11}$ state close to $\ket{02}$ state following a fast adiabatic trajectory, generating a $\pi$ phase shift on the $\ket{11}$ state~\cite{aCZ2014}. Specifically, we tune $Q_2$ from $5.31$~GHz to $5.085$~GHz, while keeping $Q_1$ at the idle point $4.835$~GHz all the time. The length of the CZ gate is $55$~ns, and the fidelity is {98.7}$\textrm{\%}$ determined by the quantum process tomography (QPT), which is shown in Fig.~\ref{exp_supp_4}\textbf{c} and \textbf{d}. A completely positive and trace-preserving (CPTP)~\cite{Knee2018} protocol is used to ensure the physical estimation of the $\chi$ matrix from QPT.

\section{Calibrate all qubits to working frequency}

\begin{figure}[!htb]\centering
	\includegraphics[width=1\linewidth]{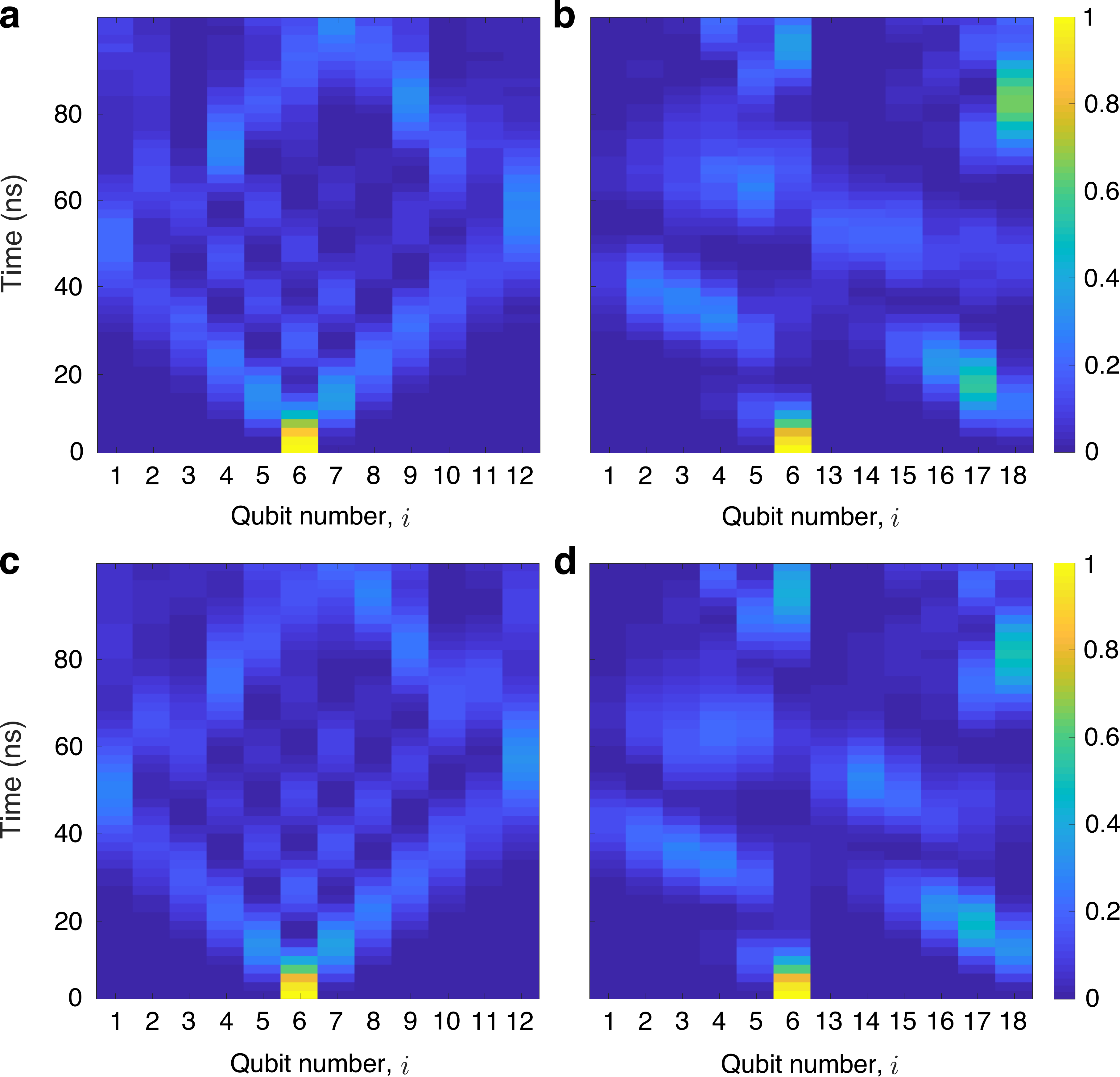}
	\caption{\textbf{Calibrate qubits to working frequency.} \textbf{a} Experimental results of the time evolution of $\langle \hat{n}_{i}\rangle$ when $Q_6$ is excited in the qubit chain. \textbf{b} Experimental results of the time evolution of $\langle \hat{n}_{i}\rangle$ when $Q_6$ is excited in the qubit ladder. \textbf{c} The numerical simulation of the data in \textbf{a}.  \textbf{d} The numerical simulation of the data in \textbf{b}. }  \label{exp_supp_5}
\end{figure}

Adjusting all qubits to the same working frequency plays an important role in this work, as the mismatch of qubit frequencies will induce an unwanted disorder. Although the calibration of $Z$ pulse distortion and $Z$ pulse crosstalk~\cite{ladder_sqp_ye,Yan2019} have been performed, the imperfect calibration of the $Z$ pulse crosstalk still results in a drift of frequency when detuning the qubits to working points. Here, we use multi-qubit excitation propagation to calibrate and correct the frequency drift. The calibration process is listed below:
\begin{itemize}
	\item[(1)] Prepare $Q_m$ among the $N$ qubits to $\ket{1}$ and leave the others in $\ket{0}$. Then we tune all qubits to the target frequency $f_m$. After an evolution time $t$, we measure the population of all sites, i.e., $\langle \hat{n}_{i}\rangle = \langle \hat{\sigma}_{i}^{+} \hat{\sigma}_{i}^{-} \rangle$. Here, in the $12$-qubits chain case, we set $f_m$ as
	\begin{eqnarray}
	f_m(m)= f_c+s\times(m-\frac{N-1}{2}),
	\label{alignmentfreq}
	\end{eqnarray}
	where $f_c$ is the chosen working frequency $f_c=4.863$~GHz, and $s$ is the frequency difference for nearest-neighbor sites, whose value is chosen as $\pm 3$~MHz in two individual measurements. Here $N=12$ is the number of qubit in the chain. We excite different sites in sequence and prepare 12 initial states for the evolution, and finally, we can get $24$ time-dependent population distributions $Z_{\textrm{exp}}$.
	\item[(2)] We use QuTiP~\cite{QuTiP1,QuTiP2} to simulate the evolution. The Hamiltonian used in the simulation is
	\begin{eqnarray}
	\hat{H}=\hat{H}_{\text{C}}+
	\sum_{n=1}^{12}\frac{n\times s+\delta_f(n)}{2}(1-2\hat{\sigma}_{n}^{-}\hat{\sigma}_{n}^{+})
	\label{H-array}
	\end{eqnarray}
	with $\hat{\sigma}_{n}^{-}$($\hat{\sigma}_{n}^{+}$ ) as the annihilation (creation) operator of the $n$-th qubit, $\hat{H}_{\text{C}}$ as the Hamiltonian (\ref{HC}) and $\delta_f$ as the independent variables which refer to the frequency drifts in tuning qubits.
	
	For the 24 time-dependent distribution $Z_{\textrm{exp}}(\textrm{ii})$ (ii=1 to 24), the population propagations are numerically simulated with corresponding initial states and $f_m$, and then the expected values $Z_{\textrm{sim}}$ for all qubits can be obtained. The distance between the numerical and experimental results is define as $Z_{\textrm{diff}}(\textrm{ii})=(Z_{\textrm{sim}}-Z_{\textrm{exp}})^2 $, and the distance of all $24$ evolutions is $Z_{\textrm{diffall}}=\sum_{\textrm{ii}=1}^{24} Z_{\textrm{diff}}(\textrm{ii})$. By changing $\delta_f$, we use Nelder-Mead optimization algorithm to minimize the distance $Z_{\textrm{diffall}}$, and finally get a $\delta_f$ array which referred to the frequency drift.
	\item[(3)] After $\delta_f$ is obtained, we add $\delta_f$ to $f_m$ as the offset calibration to correct the drift, and then repeat the steps (1)-(3) until the absolute values of frequency drift are all small and the distance $Z_{\textrm{diffall}}$ without optimization is close to the optimized distance in the previous cycle.
\end{itemize}

Fig.~\ref{exp_supp_5}\textbf{a} shows a part of the calibration results for the qubit chain, in which $Q_6$ is excited and then evolved for about $100$~ns. After two cycles of calibration, the simulation pattern of $Q_6$ is quite similar with the experiment result (Fig.~\ref{exp_supp_5}\textbf{c}), and the final distance is $Z_{\textrm{diffall}}=10$. For the $12$-qubit chain, the relative frequency drift is smaller than that presented in Table~\ref{tab2} according to the final calibration.

We use the same method to calibrate the $12$-qubits ladder, except for the alternation of the Hamiltonian
\begin{eqnarray}
\begin{aligned}
\hat{H}  = \hat{H}_{\text{L}}+\frac{1}{2}\sum_{m\in\{1,2\}}\sum_{n=1}^{6}\Delta_{mn}(1-2\hat{\sigma}_{m,n}^{-}\hat{\sigma}_{m,n}^{+}),
\label{H-ladder}
\end{aligned}
\end{eqnarray}
where the last term involves the arrangement of alignment frequencies and $\Delta_{mn}=s[m+2(n-1)] + \delta_{f}(n+6(m-1))$. Fig.~\ref{exp_supp_5}\textbf{b} shows a part of the calibration results for the $12$-qubit ladder, in which $Q_6$ is excited. After two cycles of calibration, the simulation pattern is also similar with the experiment result (Fig.~\ref{exp_supp_5}\textbf{d}). The final distance is $Z_{\textrm{diffall}}=21$, and the relative final frequency drift is smaller than that presented in Table~\ref{tab3} for the qubit ladder.

\begin{table}[t]
	\centering
	\begin{tabular}{c |c c c c c c}
		\hline
		\hline
		Qubit number
		&$\text{Q}_{1}$ &$\text{Q}_{2}$ &$\text{Q}_{3}$ &$\text{Q}_{4}$ &$\text{Q}_{5}$ &$\text{Q}_{6}$			
		\\
		
		Final frequency draft $\delta_{f}/2\pi$~(MHz)
		&0.3	&0.2 &0.3	&0.3	&0.2	&1.5\\
		\hline
		\hline
		Qubit number
		&$\text{Q}_{7}$  &$\text{Q}_{8}$	&$\text{Q}_{9}$  &$\text{Q}_{10}$  &$\text{Q}_{11}$	&$\text{Q}_{12}$\\
		Final frequency draft $\delta_{f}/2\pi $~(MHz)
		&0.0  &0.3	&0.5	&0.3	&0.1	&2.2\\
		\hline
		\hline
	\end{tabular}	
	
	\caption{
		{\textbf{The final frequency draft in the qubit chain.}  }
	}
	\label{tab2}	
\end{table}

\begin{table}[t]
	\centering
	\begin{tabular}{c |c c c c c c}
		\hline
		\hline
		Qubit number
		&$\text{Q}_{1}$ &$\text{Q}_{2}$ &$\text{Q}_{3}$ &$\text{Q}_{4}$ &$\text{Q}_{5}$ &$\text{Q}_{6}$			
		\\
		
		Final frequency draft $\delta_{f}/2\pi$~(MHz)
		&0.0	&0.2 &0.1	&0.1	&0.5	&0.4\\
		\hline
		\hline
		Qubit number
		&$\text{Q}_{13}$  &$\text{Q}_{14}$	&$\text{Q}_{15}$  &$\text{Q}_{16}$  &$\text{Q}_{17}$	&$\text{Q}_{18}$\\
		Final frequency draft $\delta_{f}/2\pi $~(MHz)
		&0.1  &0.3	&0.7	&0.3	&0.4	&0.2\\
		\hline
		\hline
	\end{tabular}	
	
	\caption{
		{\textbf{The final frequency draft in the qubit ladder.}  }
	}
	\label{tab3}	
\end{table}

\section{Local densities of the one-dimensional Bose-Hubbard model in noninteracting case }
The Hamiltonian of the one-dimensional Bose-Hubbard model reads 
\begin{eqnarray}
\hat{H}_{\text{BH}} = J\sum_{i=1}^{L}(\hat{a}_{i}^{\dagger}\hat{a}_{i+1} + \hat{a}_{i+1}^{\dagger}\hat{a}_{i}) +\frac{U}{2}\sum_{i=1}^{L}\hat{n}_{i}(\hat{n}_{i}-1)
\label{H_BH_1d}
\end{eqnarray}
with $J$ and $U$ as the standard hopping and nonlinear interaction parameters, and $\hat{n}_{i} = \hat{a}_{i}^{\dagger}\hat{a}_{i}$ as the bosonic number operator. The Hamiltonian (\ref{H_BH_1d}) can describe a superconducting qubit chain. The qubit chain used in this work satisfies $|U|/J\simeq 19$. Since the dynamics of the local densities $n_{|0(1)\rangle}$ under the unitary evolution $\exp(-i\hat{H}_{\text{BH}} t)$ can be analytically derived when $U/J\rightarrow \infty$~\cite{BH3}, we can compare the analytical results of an ideally noninteracting model with the experimental data and show that the influence of finite $U/J$ is negligible. 

\begin{figure}[]\centering
	\includegraphics[width=1\linewidth]{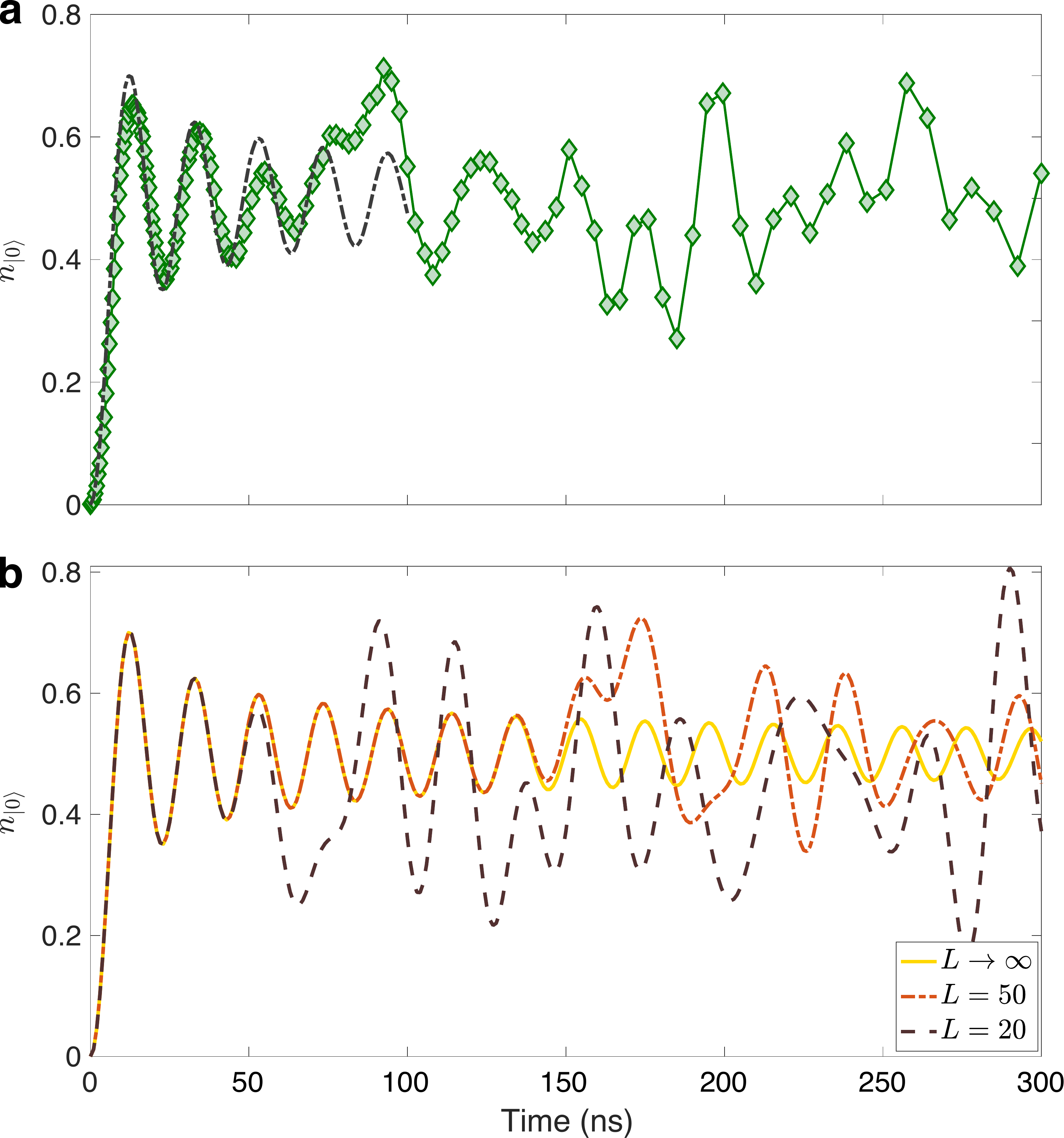}
	\caption{\textbf{Analytical results of the local densities in comparison with the experimental data.} \textbf{a} The experimental (green diamonds) and analytical (dashed line) results of the local density $n_{|0\rangle}$.  \textbf{b} The analytical results according to Eq.~\ref{local_density_0} with different system size.}\label{num_supp_1}
\end{figure}

When $U/J\rightarrow \infty$, the system reaches the hard-core limit of the Bose-Hubbard model, which can be mapped to a free-fermionic spinless model via the Jordan-Wigner transformation
\begin{eqnarray}
\hat{a}_{n} = \exp(-i\pi \sum_{m<n} \hat{f}_{m}^{\dagger}\hat{f}_{m}) \hat{f}_{n}
\label{JW}
\end{eqnarray}
where $\hat{f}_{n}$ ($\hat{f}^{\dagger}_{n}$) refers to the fermionic annihilation (creation) operator. Using a Fourier transformation, we can obtain 
\begin{eqnarray}
\hat{H}_{\text{BH}}^{U/J\rightarrow\infty} = \sum_{k=1}^{L}\lambda_{k}\hat{d}_{k}^{\dagger}\hat{d}_{k}
\label{diag_BH}
\end{eqnarray}
with $\hat{d}_{k}$ ($\hat{d}^{\dagger}_{k}$) as the fermionic operator in the momentum space, and 
\begin{eqnarray}
\lambda_{k} = 2J\cos(2\pi k/L),
\label{es}
\end{eqnarray}
where $k=1,2,...,L$. With a initial state $|\psi(0)\rangle=|1,0,1,0,...,1,0\rangle = \hat{f}_{1}\hat{f}_{3}\hat{f}_{5}... |0\rangle$ and a diagonal Hamiltonian (\ref{diag_BH}), we can directly calculate the local densities as ~\cite{BH3}
\begin{eqnarray}
n_{|1\rangle}(t) = \frac{1}{2} + \frac{1}{2L}\sum_{k=1}^{L} \exp[-4itJ\cos(2\pi k/L)],
\label{local_density_1}
\end{eqnarray}
and 
\begin{eqnarray}
n_{|0\rangle}(t) = \frac{1}{2} - \frac{1}{2L}\sum_{k=1}^{L} \exp[-4itJ\cos(2\pi k/L)].
\label{local_density_0}
\end{eqnarray}
When $L\rightarrow \infty$, the term $\frac{1}{L}\sum_{k=1}^{L} \exp[-4itJ\cos(2\pi k/L)]$ can be rewritten as a Bessel function $J_{0}(-4Jt)$. The experimental data of $n_{|0\rangle}$ and the analytical results according to Eq.~\ref{local_density_0} are plotted in Fig.~\ref{num_supp_1}\textbf{a}, showing that the short-time behavior of the experimental data
is consistent with the analytical results. The oscillation of the experimental data at later time can be regarded as a finite-size effect since the fluctuation of $n_{|0\rangle}$ becomes stronger in smaller system (see Fig.~\ref{num_supp_1}\textbf{b}). 

\section{The temperature in the Boltzmann density operator} 

Ergodic dynamics suggests that for a subsystem $A$ in the long-time and large scale limit~\cite{ergodic1}, 
\begin{eqnarray}
\rho_{A}(t) = \rho_{A}^{\text{eq}}(T), 
\label{ergodic}
\end{eqnarray}
where $\rho_{A}(t)=\text{Tr}_{B}[\rho(t)]$, $\rho_{A}^{\text{eq}}(T) = \text{Tr}_{B}[\rho^{\text{eq}}(T)]$, and $\rho(t)$ and $\rho^{\text{eq}}(T)$ refer to the quenched state at time $t$ and Boltzmann density operator with temperature $T$, respectively. According to Eq. (\ref{ergodic}), the distance $d(\rho_{A}(t),\rho^{\text{eq}}_{A}(T))$ can characterize the ergodicity. 

The temperature $T$ in Eq. (\ref{ergodic}) can be determined by the initial state, which satisfies~\cite{cold_atom,weak_and_strong}
\begin{eqnarray}
\text{Tr}\{[\rho^{\text{eq}}(T) - |\psi_{0}\rangle\langle\psi_{0}|]\hat{H}\} = 0, 
\label{initial_T}
\end{eqnarray}
where $|\psi_{0}\rangle$ is the initial state, and $\hat{H}$ is the Hamiltonian of the qubit chain or ladder. For the chosen initial states in the qubit chain and ladder, it can be directly calculated that $\langle\psi_{0}|\hat{H}|\psi_{0}\rangle\simeq 0$ and thus $1/T=0$ satisfies Eq.(\ref{initial_T}). For the following experimental and numerical results and the results in the main text, we consider the distance $d(\rho_{A}(t),\rho^{\text{eq}}_{A}(T))$ with $T\rightarrow \infty$.

\section{The impact of decoherence on the entanglement entropy}

\begin{figure}[]\centering
	\includegraphics[width=1\linewidth]{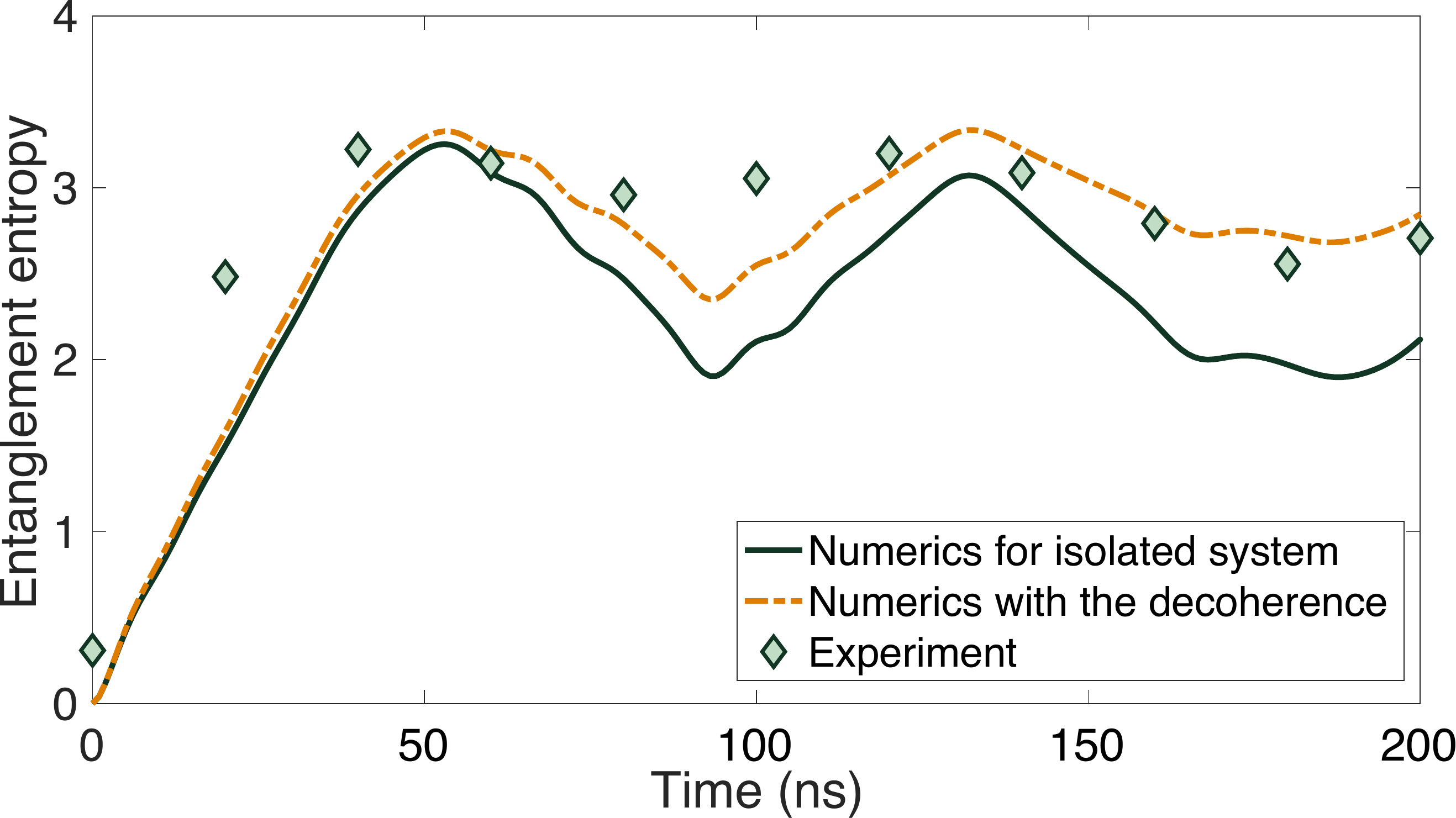}
	\caption{\textbf{Experimental and numerical results of the entanglement entropy.} The time evolution of entanglement entropy in the qubit chain with subsystem length $l=6$. The green diamonds are the experimental data. The solid line is the numerical result for an isolated system, and the dashed line is the numerical result taking the dephasing effect into consideration. }\label{num_supp_3}
\end{figure}

In the main text, the presented numerical data are calculated by considering the unitary evolution of the superconducting qubits as an isolated system. However, the coupling of the qubits to the environment is unavoidable, leading to the decoherence that may affects the dynamics of entanglement entropy. 

To quantitative estimate the effect of decoherence, we can solve the Lindblad master equation for the reduced density matrix obtained from partially tracing the environment, i.e., 
\begin{eqnarray}
\dot{\rho}(t) = &-& i[\hat{H}, \rho(t)] \\ \nonumber
&+&\frac{1}{2}\sum_{n}[2\hat{C}_{n}\rho(t)\hat{C}_{n}^{\dagger} - \{\hat{C}_{n}^{\dagger}\hat{C}_{n},\rho(t)\}]
\label{rho_decoherence}
\end{eqnarray}
with $\hat{C}_{n} = \sqrt{\gamma_{n}}\hat{A}_{n}$ as the collapse operators. There are two effects of the decoherence, i.e., the energy relaxation effect and the dephasing effect, characterized by the $T_{1}$ and $T_{2}^{*}$ in Table~\ref{tab1}, respectively. Since $T_{2}^{*} \ll T_{1}$, it is predicted that the dephasing effect is stronger than the energy relaxation effect. Hence, we can study the dephasing effect to explain the discrepancy between the experimental
and numerical results in Fig. 3\textbf{e}. The collapse operator of the dephasing effect is $\hat{C}_{n} = \hat{\sigma}^{z}_{n}/\sqrt{2T_{2,n}^{*}}$, where $T_{2,n}^{*}$ refers to the dephasing time of the $n$-th qubit. With Eq.~(\ref{rho_decoherence}), we can numerically simulate the time evolution of entanglement entropy with the dephasing effect. As shown in Fig.~\ref{num_supp_3}, the numerics considering the dephasing effect have a better agreement with the experimental data.

\begin{figure*}[!htb]\centering
	\includegraphics[width=1\linewidth]{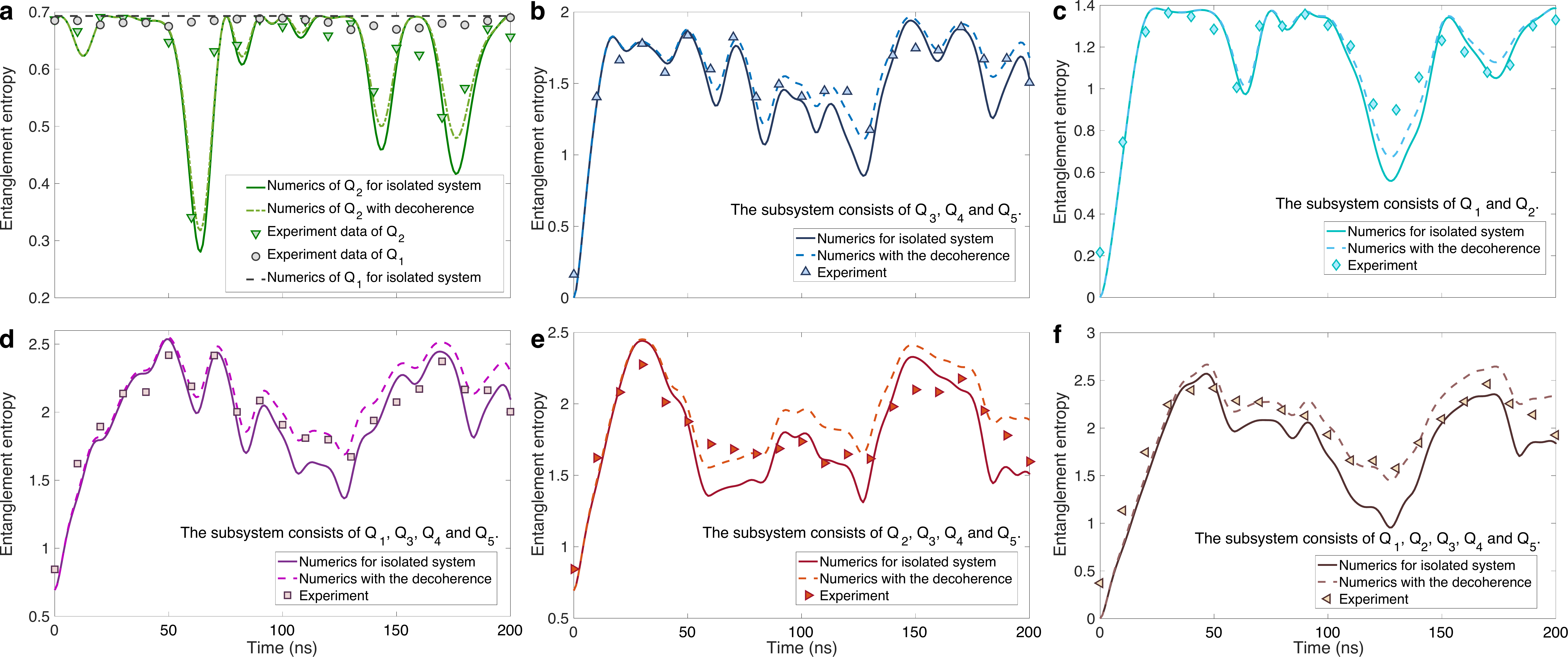}
	\caption{\textbf{Experimental and numerical results of the entanglement entropy for different subsystems in the qubit chain.} \textbf{a} The dynamics of $S(\rho_{A})$ ($Q_{1}$) and $S(\rho_{B})$ ($Q_{2}$). \textbf{b} The dynamics of $S(\rho_{C})$.  \textbf{c} The dynamics of $S(\rho_{AB})$.  \textbf{d} The dynamics of $S(\rho_{AC})$. \textbf{e} The dynamics of $S(\rho_{BC})$. \textbf{f} The dynamics of $S(\rho_{ABC})$. }  \label{num_supp_4}
\end{figure*}

\begin{figure*}[!htb]\centering
	\includegraphics[width=1\linewidth]{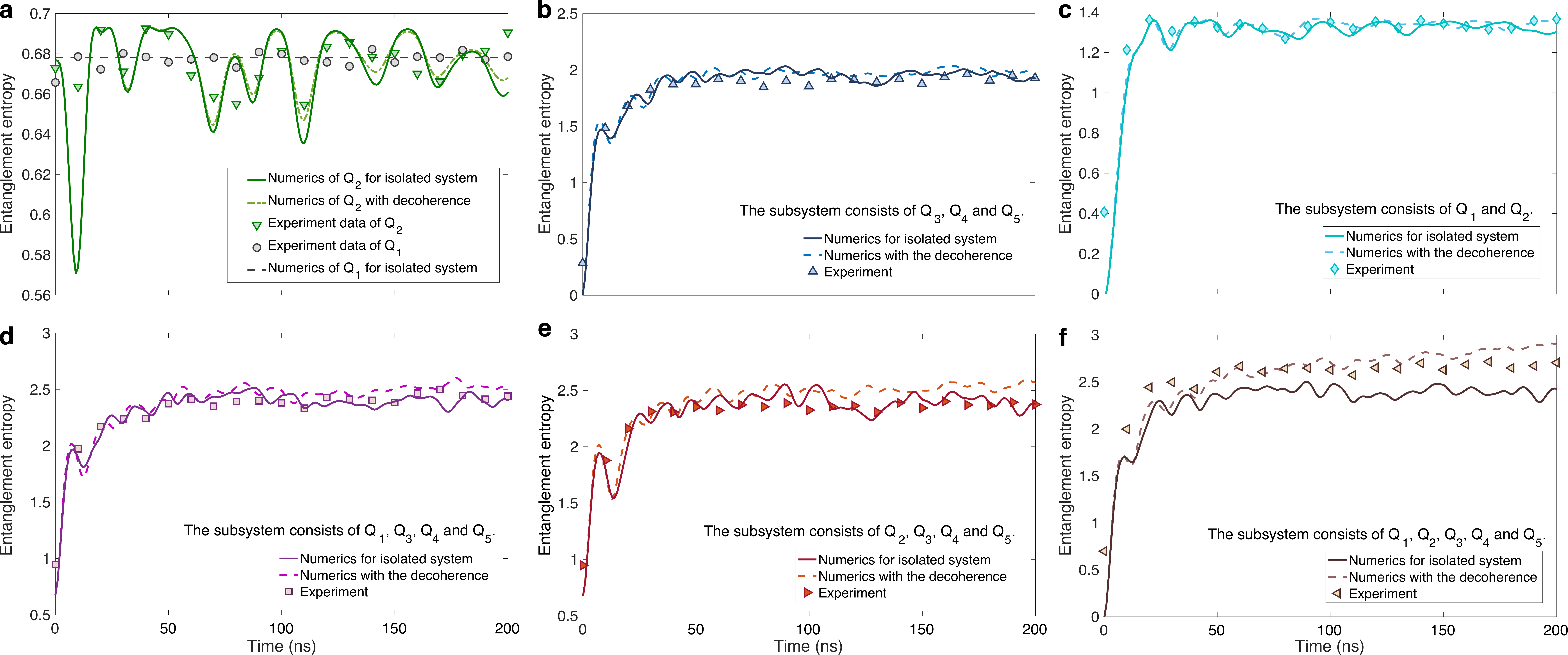}
	\caption{\textbf{Experimental and numerical results of the entanglement entropy for different subsystems in the qubit ladder.} \textbf{a} The dynamics of $S(\rho_{A})$ ($Q_{1}$) and $S(\rho_{B})$ ($Q_{2}$). \textbf{b} The dynamics of $S(\rho_{C})$.  \textbf{c} The dynamics of $S(\rho_{AB})$.  \textbf{d} The dynamics of $S(\rho_{AC})$. \textbf{e} The dynamics of $S(\rho_{BC})$. \textbf{f} The dynamics of $S(\rho_{ABC})$. }\label{num_supp_5}
\end{figure*}

\begin{figure*}[!htb]\centering
	\includegraphics[width=1\linewidth]{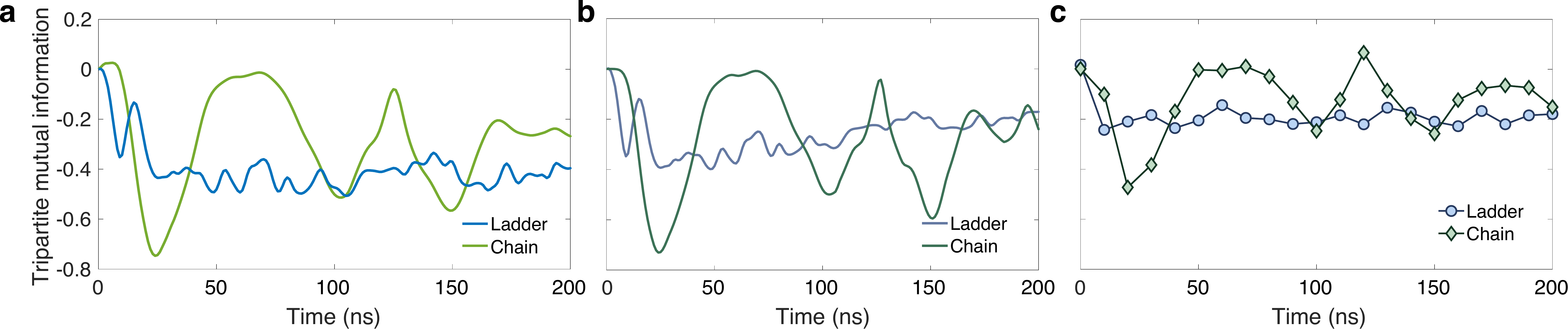}
	\caption{\textbf{Experimental and numerical results of tripartite mutual information (TMI) in the qubit chain and ladder.} \textbf{a} The numerical results of the TMI in the qubit chain and ladder without considering decoherence.  \textbf{b} The numerical results of the TMI in the qubit chain and ladder with the dephasing effect. \textbf{c} The experimental data of the TMI in the qubit chain and ladder.}\label{num_supp_6}
\end{figure*} 

\section{Numerical results of the tripartite mutual information}

In this section, we present more numerical details of the tripartite mutual information (TMI). The definition of TMI is 
\begin{eqnarray}
I_{3} &=& S(\rho_{A}) + S(\rho_{B}) + S(\rho_{C}) + S(\rho_{ABC}) \\ \nonumber &-& S(\rho_{AB}) - S(\rho_{AC}) -S(\rho_{BC}),
\label{TMI_def}
\end{eqnarray}
which actually consists of the von Neumann entropy of different subsystems. Below, we will show the dynamics of entanglement entropy for different subsystems. For both the qubit chain and ladder, the subsystem $A$ and $B$ is chosen as $Q_{1}$ and $Q_{2}$ respectively, and the subsystem $C$ is comprised of the qubit $Q_{3}$, $Q_{4}$ and $Q_{5}$. 

The results are plotted in Fig.~\ref{num_supp_4} and~\ref{num_supp_5}. The finite value of $S(\rho_{B})$ at the initial time $t=0$ indicates that the information about the qubit $Q_{1}$ is locally encoded in the qubit $Q_{2}$ through entanglement by the CNOT gate (Fig.~\ref{num_supp_4}\textbf{a} and~\ref{num_supp_5}\textbf{a}). Moreover, in Fig.~\ref{num_supp_4} and~\ref{num_supp_5}, it is seen that although the values of entanglement entropy are influenced by the decoherence, the overall non-equilibrium behaviors of entanglement entropy does not significantly affected by the decoherence. Consequently, the experimental data of TMI comprised of the results in Fig.~\ref{num_supp_4} and~\ref{num_supp_5} can reveal the distinct difference between the information scrambling in the qubit chain and ladder.  
We then present the numerical results of the TMI in comparison with the experimental data (Fig.~\ref{num_supp_6}).

\end{document}